\documentclass[preprint2]{aastex631}

\usepackage[T1]{fontenc}		
\usepackage{enumitem}           

\newcounter{species}
\def\ion#1#2{\setcounter{species}{#2}#1$\;${\sc\roman{species}}\relax}

\def\l{$\lambda$}

\newcommand{\kms}{km~s$^{-1}$}

\def\arcsec{\hbox{$^{\prime\prime}$}}
\newcommand{\Msun}{M$_{\odot}$}

\newcommand{\CIVdblt}{C\,{\sc iv}$\,\lambda\lambda 1548,\!1550$}

\newcommand{\OIIIdblt}{[O\,{\sc iii}]~$\lambda\lambda 4959,\!5007$}

\newcommand{\OIdblt}{[O\,{\sc i}]~$\lambda\lambda 6300,\!6363$}
\newcommand{\SIIdblt}{[S\,{\sc ii}]~$\lambda\lambda 6716,\!6731$}
\newcommand{\NIIdblt}{[N\,{\sc ii}]~$\lambda\lambda 6548,\!6583$}

\newcommand{\CIV}{C\,{\sc iv}}

\newcommand{\Lya}{Ly$\alpha$}

\newcommand{\Hb}{{H}$\beta$}
\newcommand{\Ha}{{H}$\alpha$}
\newcommand{\FeII}{Fe\,{\sc ii}}

\newcommand{\MgII}{Mg\,{\sc ii}}

\newcommand{\OIII}{[O\,{\sc iii}]}

\def\lsim{\lower0.3em\hbox{$\,\buildrel <\over\sim\,$}}
\def\gsim{\lower0.3em\hbox{$\,\buildrel >\over\sim\,$}}

\newcommand{\pone}{Paper~\textsc{I}}

\def\kms{\,km~s$^{-1}$}      

\def\sun{\hbox{$\odot$}}
\def\earth{\hbox{$\oplus$}}
\def\lesssim{\mathrel{\hbox{\rlap{\hbox{%
 \lower4pt\hbox{$\sim$}}}\hbox{$<$}}}}
\def\gtrsim{\mathrel{\hbox{\rlap{\hbox{%
 \lower4pt\hbox{$\sim$}}}\hbox{$>$}}}}

\def\arcsec{\hbox{$^{\prime\prime}$}}

\setlist[itemize]{leftmargin=0em} 

\begin{document}
\title[UV spectroscopy of supermassive black hole binaries]{A Large Systematic Search for Close Supermassive Binary and Rapidly Recoiling Black Holes - IV. Ultraviolet spectroscopy 
\footnote{Based on observations made with the NASA/ESA Hubble Space Telescope, obtained at the Space Telescope Science Institute, which is operated by the Association of Universities for Research in Astronomy, Inc., under NASA contract NAS5-26555. These observations are associated with program GO-12299.}
\footnote{Based on observations obtained with the Hobby-Eberly Telescope (HET), which is a joint project of the University of Texas at Austin, the Pennsylvania State University, Ludwig-Maximillians-Universitaet Muenchen, and Georg-August Universitaet Goettingen. The HET is named in honor of its principal benefactors, William P. Hobby and Robert E. Eberly.} 
}

\author[0000-0001-8557-2822]{Jessie C. Runnoe}
\affiliation{Vanderbilt University, Department of Physics \& Astronomy, 6301 Stevenson Center, Nashville, TN 7235, USA}
\affiliation{Fisk University, Department of Life and Physical Sciences, 1000 17th Avenue N, Nashville, TN 37208, USA}

\author[0000-0002-3719-940X]{Michael Eracleous}
\affiliation{Department of Astronomy \& Astrophysics, and Institute for Gravitation and the Cosmos, Penn State University, 525 Davey Lab, 251 Pollock Road, University Park, PA 16802, USA}


\author[0000-0002-7835-7814]{Tamara Bogdanovi{\'c}}
\affiliation{Center for Relativistic Astrophysics, School of Physics, Georgia Institute of Technology, Atlanta, GA 30332}

\author[0000-0003-4814-2377]{Jules P. Halpern}
\affiliation{Columbia Astrophysics Laboratory, Columbia University, 550 West 120th Street, New York, NY 10027-6601 }

\author[0000-0002-8187-1144]{Steinn Sigur{\dh}sson}
\affiliation{Department of Astronomy and Astrophysics, and Institute for Gravitation and the Cosmos, Penn State University, 525 Davey Lab, 251 Pollock Road, University Park, PA 16802, USA}





\begin{abstract}
We present Hubble Space Telescope ultraviolet (UV) of 13 quasars at $z<0.7$, along with contemporaneous optical spectra from ground-based telescopes. The targets were selected to have broad \Hb\ emission lines with substantial velocity offsets relative to the rest frame of their host galaxy. By analogy to single-line spectroscopic binary stars, these objects have been regarded as supermassive black hole binary (SBHB) candidates where the offset emission lines may be caused by bulk orbital motion. The best alternative explanation is that the \Hb\ line profile is the result of non-axisymmetric emission from a disk-like broad-line region associated with a single supermassive black hole. We use the broad UV line profiles to discriminate between these two scenarios. We describe our methodology for isolating the broad optical and UV line profiles and the criteria we apply for comparing them. Of the 13 SBHB candidates, three have strong evidence in support of the SBHB hypothesis, five have tentative support, one is disfavored, and four have such severely absorbed UV line profiles that the results are inconclusive.
%

\end{abstract}




\section{Introduction}
\label{sec:intro}

\subsection{Background and Motivation}
Supermassive black hole binaries (SBHBs) are expected to be a product of hierarchical galaxy evolution \citep{white78,white91} given that all massive galaxies host central supermassive black holes \citep{kormendy93,kormendy95,kormendy13}. Despite examples of kiloparsec-scale dual active galactic nuclei \citep[AGN; e.g.,][]{comerford09b,comerford15,foord20,chen22}, the close gravitationally bound binaries that are expected to follow have proven elusive. The best support for their existence comes in the form of recent evidence for a low-frequency gravitational wave background \citep{nanograv_15yr_gwb,reardon23,cpta23,epta24} found with Pulsar Timing Arrays \citep[PTAs;][]{mclaughlin13,hobbs13,kramer13}; at their closest separations, SBHBs become loud gravitational wave sources and the observations are consistent with the stochastic background produced by a cosmological population of SBHBs \citep{nanograv_15yr_astrophys}.

There are a number of electromagnetic signatures that have been used to search for SBHBs at different separations \citep[for a review, see][and references therein]{bogdanovic22,derosa19,dorazio23}, typically under the assumption that gas is present and at least one member of the SBHB is accreting as an AGN. One such method for identifying SBHB candidates is to make analogy to spectroscopic binary stars and look for periodic Doppler shifts in quasar broad emission lines relative to their narrow lines \citep{gaskell83, halpern88}. This approach is sensitive to SBHBs with sub-parsec separations and periods of order decades to hundreds of years \citep{eracleous12,pflueger18,kelley21}. 

This radial velocity technique for finding and evaluating SBHB candidates was applied early on to a class of AGN with double peaked broad emission lines \citep{halpern88,gaskell96a}. In the most pronounced cases, these objects have extremely broad, double peaked Balmer lines in their rest-frame optical spectra \citep{stauffer83,halpern90,storchi-bergmann93,eracleous94,eracleous03,eracleous09}, with less pronounced cases corresponding to very boxy or flat-topped line profiles \citep{strateva03}. Interpreted in the context of SBHBs, each peak in a Balmer line would be attributable to one member of the binary and would be expected to shift in velocity over the course of the orbit. However, for many reasons, the SBHB explanation for such objects is no longer preferred \citep[see discussion in][]{eracleous97,liu16,doan20}. Instead, the double-peaked broad-line profiles are consistent with a disk-like broad-line region (BLR) around a single supermassive black hole \citep{eracleous94,eracleous03,storchi-bergmann17,schimoia17} and are therefore commonly referred to as ``disk-like emitters'' (DEs). The same model for the BLR can also produce the single-peaked Balmer profiles observed in the majority of quasars for particular disk parameters \citep{jackson91,eracleous03,nguyen16}, a region that produces a broad Gaussian profile that fills in the gaps between the peaks \citep{lewis10}, or radiative transfer effects as in a disk wind \citep{murray97,flohic12,nguyen19}. Emissivity perturbations \citep[e.g., from an eccentric disk or spiral arm;][]{eracleous95,storchi-bergmann97,storchi-bergmann03,schimoia12,schimoia17} that change on dynamical timescales in the BLR can produce variation, including systematic drift, in the velocity of the broad-line peaks on timescales of years \citep{lewis10} that mimics the trend of orbital motion in a binary.

\subsection{Ultraviolet spectroscopy as a test of binary supermassive black hole candidates}
Ultraviolet (UV) spectroscopy played a critical role in testing the SBHB hypothesis for DEs. The discriminatory power of the UV lines comes from the fact that bulk orbital motion in an SBHB will produce the same Doppler shift in all emission lines regardless of wavelength. So, if this is not observed, the SBHB hypothesis can be ruled out. It is more difficult to interpret the scenario where the UV and optical lines have matching velocity offsets because this can arise from both single black holes and SBHBs.

Expectations for comparing the optical and UV broad lines of single black holes is rooted in DEs. Notably, the UV broad emission lines of DEs differ substantially from the broad Balmer profiles \citep{halpern96,eracleous06,eracleous09}. They can be asymmetric or have a broad base and narrow core and are often narrower overall. There is a pronounced difference between the Balmer lines and Ly$\alpha$ and \CIVdblt. Even though the Balmer lines show double-peaked, boxy profiles, Ly$\alpha$ and, often, \ion{C}{4} are single-peaked with cuspy profiles, narrower than the Balmer lines at half maximum, and have no velocity offset. The difference in profiles has been attributed to a combination of collisional destruction of Ly$\alpha$ and \ion{C}{4} photons in the dense inner disk \citep{collin-souffrin90} and radiative transfer in the rapidly-accelerating base of a disk wind \citep{murray97,flohic12,chajet13}. Thus, unlike Balmer photons, the Ly$\alpha$ and \ion{C}{4} photons are either destroyed in the disk or diffuse in an outflowing layer above it, so the profiles of these lines lose the double-peaked shape characteristic of disk emission \citep{eracleous09}. 

This picture is continued in more recent, numerical models of UV resonance line formation in an outflowing wind above the accretion disk \citep[e.g.,][and references therein]{matthews20,matthews23} that can produce both single-peaked and double-peaked profiles for the UV resonance lines. The shape of the line profiles depends sensitively on the velocity law of the wind, on other parameters like the viewing angle, and on the importance of resonance scattering. Winds that accelerate slowly have their line formation regions in their outer parts, where the outflow velocity exceeds the rotational velocity and produce single-peaked UV line profiles. Vice versa, winds that accelerate quickly produce profiles that are boxy or double-peaked. As a corollary, lines produced in parts of the wind where the rotation speed dominates over the outflow speed (e.g., the Balmer lines produced at the base of the wind, at low latitudes) will be double peaked while lines produced in the parts of the wind where the velocity along the radial streamlines dominates (e.g., the UV resonance lines produced at high latitudes) will be single peaked.

Although DEs do not appear to be SBHBs, the single-black-hole-active scenario remains an open parameter space to search for candidates. Here, candidates are AGN selected based on the shifts of their broad emission lines relative to the rest-frame set by narrow [\ion{O}{3}]~$\lambda 5007$. One possibility, corresponding to SBHBs in conjunction, is to select single-peaked broad \Hb\ emission lines at the rest frame which show time-dependent velocity shifts \citep{shen13a}. Another is where the broad \ion{Mg}{2} \citep{ju13} or \Hb\ emission lines \citep{tsalmantza11,eracleous12,liu14} have a large velocity offset relative to the rest frame. In this Doppler shift scenario, the shifted broad lines are associated with the active member of an SBHB where the velocity offset and subsequent shifts \citep{wang17,runnoe17,guo19} are the result of bulk orbital motion of the active black hole and its bound BLR. The SBHB candidates selected via this approach are consistent with SBHBs having total masses of order $10^8$~\Msun, mass ratios of order $10^{-1}$, 0.01--0.1~pc separations, and periods of order decades to a few hundred years \citep{eracleous12,runnoe15,runnoe17,pflueger18,guo19,nguyen20a,kelley21}. 

However, the SBHB interpretation is not a unique explanation for velocity offsets and shifts in the broad emission lines of AGN \citep[e.g.,][]{bogdanovic09}. The best alternative can be found by analogy to DEs: the velocity-offset, single-peaked broad Balmer lines and their velocity shifts in repeat spectroscopy are produced by a perturbed accretion disk around a single supermassive black hole and its dynamical evolution. In this scenario, the offset \Hb\ peak observed in single-peaked SBHB candidates is effectively one of the two peaks of a double-peaked line significantly strengthened by non-axisymmetric structures in the disk \citep{shields09,chornock10,lauer09}. The peaks of the broad lines then drift in velocity on the BLR dynamical time, which is of order years for a typical AGN. Examples of broad \Ha\ profiles that appear double-peaked are some epochs but vary with time so that they appear to have one pronounced, shifted peak at other epochs can be found in \citet[][see 3C~390.3 in their Fig.~42 and PKS~0235+023 in their Fig.~45]{gezari07}, \citet[][see 1E~$0450.3-1817$ in their Fig.~3 and Pictor~A in their Fig.~4]{lewis10}, and \citet[][see their Fig.~2]{schimoia17}.

The purpose of this paper, the fourth in a series, is to test the best alternative scenario for explaining the radial velocity measurements in some of the best SBHB candidates with single-peaked velocity offset optical broad lines: that they are produced by a perturbed accretion disk, perhaps with an associated wind, around a single black hole. The test uses the broad UV resonance lines (especially \Lya) and optical Balmer lines and relies on the models of line formation described above. In summary, if a candidate SBHB with shifted Balmer lines shows different shifts in its UV lines, and especially if the UV lines have very small shifts, then it is disfavored as a candidate. If, however, a candidate SBHB shows UV lines with similar shifts as the Balmer lines, then it passes the test and it becomes a stronger candidate. But passing the test does not conclusively establish a candidate as a true SBHB.

This paper is organized as follows. In Section~\ref{sec:data} we describe how the sample was identified and new observations taken to support the test. The spectral decomposition results are presented in Section~\ref{sec:sfit} and the analysis, methods for comparing the optical and UV spectra, and notes on individual objects, are presented in Section~\ref{sec:analysis}. The outcomes of this work are discussed in Section~\ref{sec:discussion} and summarized in Section~\ref{sec:summary}.

\section{Sample and Data}
\label{sec:data}
The initial sample of SBHB candidates in this work was selected by \citet[][;\pone\ in this series]{eracleous12} from among 15,900 $z<0.7$ quasars \citep{schneider10} in the seventh data release \citep[DR7;][]{sdssdr7} of the Sloan Digital Sky Survey \citep[SDSS;][]{york00,richards02,gunn06,smee13}. Principal component analysis \citep{boroson10} identified sources with $|\Delta v|\gtrsim1000$~\kms\ velocity offsets between the broad \Hb\ emission line and the rest frame set by [\ion{O}{3}]~$\lambda 5007$ and visual inspection further required that the broad lines be single peaked. The resulting sample included 88 candidates whose flux variability \citep{runnoe15} and radial velocity curves \citep{runnoe17} were presented in Papers~\textsc{II} and \textsc{III} in this series, respectively.

A subsample of 13 of the SBHB candidates were targeted for UV spectroscopy with the HST. These were selected based on two considerations. First, we required $V<18$ so that spectra of \Lya\ and \MgII\ could be obtained in a single HST orbit. Second, we identified objects with \Hb\ profiles clearly showing shifted peaks with a variety of velocity offsets and profile asymmetries (e.g., sharp versus round peaks, symmetric versus skewed profiles, and a balance between redshifts and blueshifts). The final HST subsample consists of quasars with $0.12 < z < 0.58$ and $16.6<V<17.9$, thus covering approximately a factor of 30 in luminosity. The broad \Hb\ velocity offsets are in the range 700~km~s$^{-1}\lesssim |\Delta v|\lesssim4000$~km~s$^{-1}$. The targets and their redshifts are listed in Table~\ref{tab:obslog}, with additional object information available from \citet{eracleous12}. Table~\ref{tab:obslog} we give the the full coordinate designation of each object while elsewhere in the paper we refer to the objects with the first six digits of their right ascension, following the convention of \citet{eracleous12}.

The UV spectra that we present were obtained with the HST in 2010 and 2011 as part of the GO-12299 program (P.I. Eracleous). The \Lya\ line was observed for all 13 objects with the Cosmic Origins Spectrograph (COS) using the G140L and G230L gratings, depending on redshift, and the resulting spectra were reduced using the standard {\tt calcos} pipeline \citep{fox15}. The final COS data products were downloaded in July 2016, after the correction of a data calibration issue that introduced up to 1~\AA\ mis-registration in G230L spectra. 

In eight lower-redshift sources, we also used the Space Telescope Imaging Spectrograph (STIS) with the G430L grating to observe the \MgII\ emission line. These spectra were reduced using the standard {\tt calstis} pipeline \citep{bostroem11}. We retrieved three reduced spectra per object, the result of a 3-point dither pattern, and combined them in order to improve the signal-to-noise (S/N) ratio. To do this, the spectra were first rebinned to a common wavelength scale, while preserving flux, and then averaged. The final flux in each pixel is a linear combination of the input fluxes, so the flux uncertainties were straightforward to propagate. The resulting spectra included a number of hot pixels that were not successfully removed by the pipeline. The flux values in these pixels were excised and replaced by interpolating between the two neighboring pixels, but their uncertainties were preserved.

\begin{figure*}
\centerline{
    \includegraphics[width=0.75\textwidth]{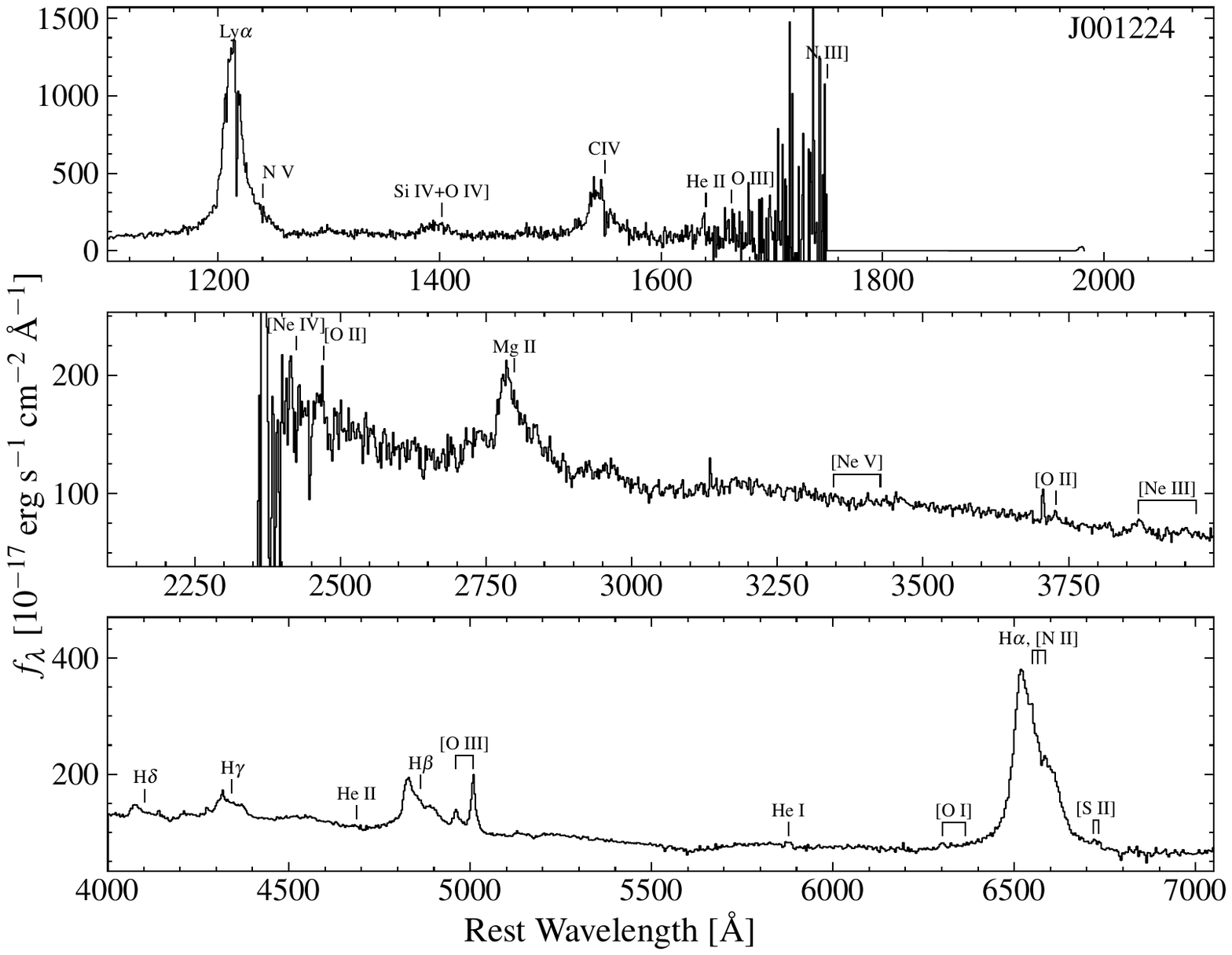} 
}
\caption{The optical/UV spectra of the SBHB candidates with HST coverage. Spectral coverage of the prominent broad emission lines was obtained contemporaneously from the telescopes and instruments described in Section~\ref{sec:data} and summarized in Table~\ref{tab:obslog}. Black lines and labels mark the nominal wavelengths of some broad and narrow emission lines common in quasar spectra \citep[e.g.,][]{vandenberk01}. When present, the \Ha\ spectra are scaled to match the \Hb\ spectrum in the optical continuum between the Balmer lines. The spectra are presented without a Galactic extinction correction.}
\label{fig:fullspec}
\end{figure*}

Finally, in the case of both the COS and STIS spectra, we sometimes binned the spectra in order to trade unnecessary spectral resolution for S/N. Note that wavelengths are on a vacuum wavelength scale and the spectra are presented and used here without Galactic extinction corrections. The details of the instrument setup and information about the final spectra are listed in Table~\ref{tab:obslog} and the spectra are presented in Figure~\ref{fig:fullspec}. Additional details of the ground-based instrument configurations and data reduction steps are described in \citet{eracleous12} and \citet{runnoe15}.

In order to facilitate a comparison between optical and UV line profiles, we obtained ground-based observations of the \Hb, \Ha, and, when appropriate, the \MgII\ profiles contemporaneously with the HST observations. The maximum time interval between the acquisition of spectra covering the \Lya\ and \Hb\ emission lines is 6.5~months (J171448), although most of the spectra were taken within a month of each other. The majority of the optical spectra were presented in \citet{runnoe15}, although a few were not included in that work due to data issues (e.g., incomplete wavelength coverage of an emission line) and are presented here for the first time. All the data reduction and calibration procedures are described in \cite{runnoe15}, including the telescopes and instruments used. Detailed information of the optical spectra is also presented in Table~\ref{tab:obslog}. 

\startlongtable
\begin{deluxetable*}{rcccccccc}
\setlength{\tabcolsep}{4pt}
\tablecolumns{9}
\tablewidth{\textwidth}
\tablecaption{Observation Log
\label{tab:obslog}
}
\tablehead{
\colhead{Object} & 
\colhead{Broad} & 
\colhead{Observation} & 
\colhead{} & 
\colhead{Exp.} & 
\colhead{} & 
\colhead{Rest-frame} & 
\colhead{S/N and} \\
\colhead{and Redshift} & 
\colhead{Emission} & 
\colhead{Date} & 
\colhead{Instr.} & 
\colhead{Time} & 
\colhead{Aperture} & 
\colhead{Wavelength} & 
\colhead{Dispersion} \\
\colhead{(SDSS J)} & 
\colhead{Line(s)\tablenotemark{b}} & 
\colhead{(UT)} & 
\colhead{Code\tablenotemark{a}} & 
\colhead{(s)} &  
\colhead{(\arcsec)} & 
\colhead{Range (\AA)} & 
\colhead{\textrm{(\AA\ pix$^{-1}$)}\tablenotemark{c}} 
}
\startdata
001224.03$-$102226.2 & \Lya, \CIV           & 2010 Oct 23 & COS  & 1996 & 2.5 \tablenotemark{d}             & $1030-1982$ & \phantom{1}6\,/\,0.7  \\
$z=0.2287$           & \MgII 		        & 2011 May 14 & STIS & 2184 & $0.35\times0.2$\tablenotemark{d}  & $2357-4645$ & 21\,/\,2.2  \\
                     & \Hb 		           & 2010 Oct 12 & HET  & 1798 & 1.5 				               & $3481-5911$ & 72\,/\,1.6  \\
                     & \Ha 		           & 2014 Aug 28 & APO  & 1560 & 1.5 				               & $4196-8012$ & 18\,/\,1.9  \\
\noalign{\vskip 8pt}                 
015530.02$-$085704.0 & \Lya, \CIV           & 2010 Oct 07 & COS  & 2088 & 2.5 \tablenotemark{d} 	        & $1087-2090$ & \phantom{1}9\,/\,0.7  \\
$z=0.1648$           & \MgII 		        & 2011 Feb 02 & STIS & 2184 & $0.35\times0.2$\tablenotemark{d}  & $2486-4897$ & 13\,/\,2.4  \\
                     & \Hb\tablenotemark{e} & 2010 Oct 02 & HET  & 1800 & 1.5 			                   & $4276-7261$ & 39\,/\,2.0  \\
                     & \Ha 		           & 2014 Aug 28 & APO  & 1260 & 1.5  				               & $4426-8451$ & 49\,/\,2.0  \\
\noalign{\vskip 8pt}                 
093653.85+533126.9 & \Lya, \CIV 	        & 2010 Oct 07 & COS  & 2384 & 2.5 \tablenotemark{d} 	        & $1031-1749$ & \phantom{1}8\,/\,0.7  \\
$z=0.2276$         & \MgII 		            & 2011 Mar 24 & STIS & 2388 & $0.35\times0.2$\tablenotemark{d}  & $2357-4644$ & 17\,/\,2.2  \\
                   & \Hb 		             & 2010 Nov 03 & HET  & 1800 & 1.5 				                 & $3486-5915$ & 60\,/\,1.6  \\
                   & \Ha 		             & 2002 Jan 07 & SDSS & 4806 & 3.0\tablenotemark{d} 	         & $3092-7511$ & 49\,/\,1.6  \\
\noalign{\vskip 8pt}                 
094603.95+013923.6 & \Lya, \CIV  	        & 2010 Dec 05 & COS  & 2124 & 2.5 \tablenotemark{d} 	        & $1037-1759$ & \phantom{1}2\,/\,0.7  \\
$z=0.2198$         & \MgII 		            & 2011 Mar 24 & STIS & 2172 & $0.35\times0.2$\tablenotemark{d}  & $2372-4674$ & \phantom{1}5\,/\,2.3  \\
                   & \Hb 		             & 2011 Jan 02 & MDM  & 5400 & 1.2  				             & $3243-5648$ & 14\,/\,0.6  \\
                   & \Ha 		             & 2001 Mar 21 & SDSS & 8106 & 3.0\tablenotemark{d} 	         & $3114-7560$ & 23\,/\,1.6  \\
\noalign{\vskip 8pt}                 
102839.11+450009.3 & \Lya  		            & 2011 Mar 09 & COS2 & 2501 & 2.5 \tablenotemark{d} 	        & $1049-1340$ & 10\,/\,1.0  \\
$z=0.5843$         & \MgII 		            & 2003 Dec 17 & SDSS & 3840 & 3.0\tablenotemark{d} 	            & $2394-5820$ & 13\,/\,0.7  \\
                   & \Hb 		             & 2011 May 21 & HET 	& 1800 & 1.5  				             & $3917-5728$ & 36\,/\,1.2  \\
\noalign{\vskip 8pt}                 
111916.13+110107.1	 & \Lya  		        & 2011 Apr 07 & COS  & 2328 & 2.5 \tablenotemark{d} 	        & $1193-1523$ & \phantom{1}3\,/\,1.1  \\
$z=0.3936$           & \MgII 		        & 2010 May 14 & Pal  & 1350 & 1.5  				                & $2274-4029$ & 22\,/\,0.8  \\
                     & \Hb 		           & 2011 Apr 05 & HET  & 2700 & 1.5  				               & $3073-5217$ & 29\,/\,1.4  \\
                     & \Ha 		           & 2004 Feb 27 & SDSS & 3900 & 3.0\tablenotemark{d} 	           & $2718-6617$ & 25\,/\,1.4  \\
\noalign{\vskip 8pt}                 
115449.42+013443.5 & \Lya  		            & 2011 May 29 & COS2 & 2308 & 2.5 \tablenotemark{d} 	        & $1128-1441$ & \phantom{1}4\,/\,2.7  \\
$z=0.4693 $        & \MgII 		            & 2010 May 14 & Pal  & 2161 & 1.5  				                & $2158-3821$ & 23\,/\,0.7  \\
                   & \Hb 		             & 2011 May 26 & HET  & 3150 & 1.5  				             & $2918-4950$ & 41\,/\,1.4  \\
\noalign{\vskip 8pt}                 
125142.28+240435.3 & \Lya, \CIV 	        & 2011 Mar 16 & COS  & 2188 & 2.5 \tablenotemark{d} 	        & $1064-1808$ & \phantom{1}7\,/\,0.7  \\
$z=0.1887$         & \MgII 		            & 2011 Apr 26 & STIS & 2184 & $0.35\times0.2$\tablenotemark{d}  & $2435-4799$ & 25\,/\,2.3  \\
                   & \Hb	 	             & 2011 Jan 02 & MDM  & 5400 & 1.2  				             & $3328-5796$ & 31\,/\,0.6  \\
                   & \Ha 		             & 2008 Feb 09 & SDSS & 3905 & 3.0\tablenotemark{d} 	         & $3199-7757$ & 35\,/\,1.6  \\
\noalign{\vskip 8pt}                 
130534.49+181932.9 & \Lya, \CIV 	        & 2011 Jan 25 & COS  & 2068 & 2.5 \tablenotemark{d} 	        & $1133-1899$ & 16\,/\,0.7  \\
$z=0.1177$         & \MgII 		            & 2011 Apr 17 & STIS & 2190 & $0.35\times0.2$\tablenotemark{d}  & $2589-5102$ & 15\,/\,2.5  \\
                   & \Hb 		             & 2011 Jan 29 & HET  & 1800 & 1.5  				             & $3830-6496$ & 93\,/\,1.8  \\
                   & \Ha 		             & 2008 Jan 14 & SDSS & 5405 & 3.0\tablenotemark{d} 	         & $3400-8250$ & 52\,/\,1.6  \\
\noalign{\vskip 8pt}                 
134617.55+622045.4 & \Lya, \CIV 	        & 2011 May 04 & COS  & 2469 & 2.5 \tablenotemark{d} 	        & $1133-2181$ & \phantom{1}4\,/\,0.7  \\
$z=0.1165 $        & \MgII 		            & 2010 Nov 17 & STIS & 2454 & $0.35\times0.2$\tablenotemark{d}  & $2593-5109$ & \phantom{1}8\,/\,4.9  \\
                   & \Hb 		             & 2011 Jan 08 & HET  & 1800 & 1.5  				             & $3831-6505$ & 29\,/\,1.8  \\
                   & \Ha 		             & 2001 Jun 19 & SDSS & 2702 & 3.0\tablenotemark{d} 	         & $3404-8258$ & 26\,/\,1.6  \\
\noalign{\vskip 8pt}                 
153636.22+044127.0 & \Lya  		            & 2011 Jul 23 & COS  & 2312 & 2.5 \tablenotemark{d} 	        & $1196-1527$ & 7\,/\,1.7   \\
$z=0.3889$         & \MgII 		            & 2010 May 13 & Pal  & 3000 & 1.5 				                & $2281-4042$ & 49\,/\,0.8  \\
                   & \Hb 		             & 2011 Jul 20 & HET  & 2400 & 1.5 				                 & $3086-5231$ & 47\,/\,1.4  \\
                   & \Ha 		             & 2008 Apr 07 & SDSS & 4806 & 3.0\tablenotemark{d}              & $2737-6639$ & 21\,/\,1.4  \\
\noalign{\vskip 8pt}                 
154340.74+112801.3 & \Lya  		            & 2011 Feb 10 & COS2 & 2200 & 2.5 \tablenotemark{d} 	        & $1169-1491$ & 11\,/\,1.1  \\
$z=0.4214$         & \Hb 		            & 2011 Feb 05 & HET  & 2700 & 1.5  				                & $3013-5112$ & 18\,/\,1.4  \\
                   & \Ha 		             & 2014 May 27 & APO  & 1080 & 1.5  				             & $3636-6932$ & 32\,/\,1.6  \\
\noalign{\vskip 8pt}                 
171448.51+332738.3 & \Lya, \CIV 	        & 2010 Oct 10 & COS  & 2120 & 2.5 \tablenotemark{d} 	        & $1072-2061$ & \phantom{1}8\,/\,0.7  \\
$z=0.1802$         & \MgII 		            & 2010 Oct 15 & STIS & 2220 & $0.35\times0.2$\tablenotemark{d}  & $2453-4832$ & 12\,/\,2.3  \\
                   & \Hb 		             & 2010 Mar 21 & KPNO & 3000 & 1.5  				             & $4162-6628$ & 37\,/\,1.2  \\
                   & \Ha 		             & 2008 May 05 & SDSS & 5404 & 3.0\tablenotemark{d} 	         & $3222-7813$ & 22\,/\,1.6  \\	
\enddata
\tablenotetext{a}{Inst. Codes: HST COS/NUV G140L (COS); HST COS/NUV G230L (COS2); HST STIS/CCD G430L (STIS); HET LRS G2 grism (HET); APO ARC 3.5m DIS Red (APO); SDSS 2.5m SDSS spec. (SDSS); MDM Hiltner 2.4m OSMOS (MDM); Palomar 200in DBSP Blue (Pal); KPNO Mayall 4m RCspec (KPNO)}
\tablenotetext{b}{Spectrum used for broad emission line profile shown in Figure~\ref{fig:vstack}.}
\tablenotetext{c}{The S/N per pixel measured in the continuum red of the relevant emission line, followed by the spectral width of a pixel.}
\tablenotetext{d}{Diameter of a circular aperture or fiber in the case of SDSS spectra and the size of the extraction box for STIS spectra.}
\tablenotetext{e}{This optical spectrum is presented for the first time in this work.}
\end{deluxetable*}

\begin{figure*}
\centerline{
    \includegraphics[width=0.75\textwidth]{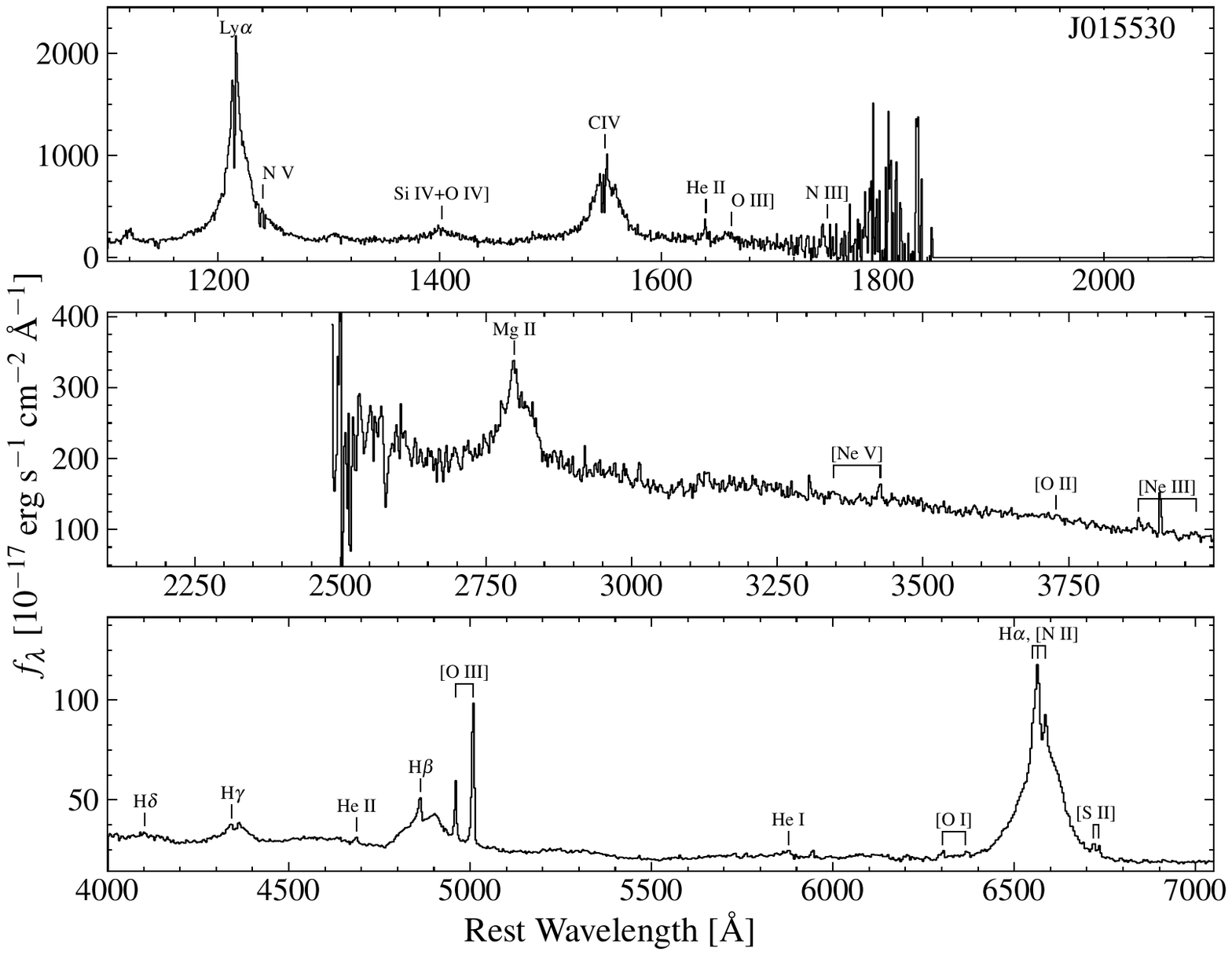} 	
}
\centerline{  
    \includegraphics[width=0.75\textwidth]{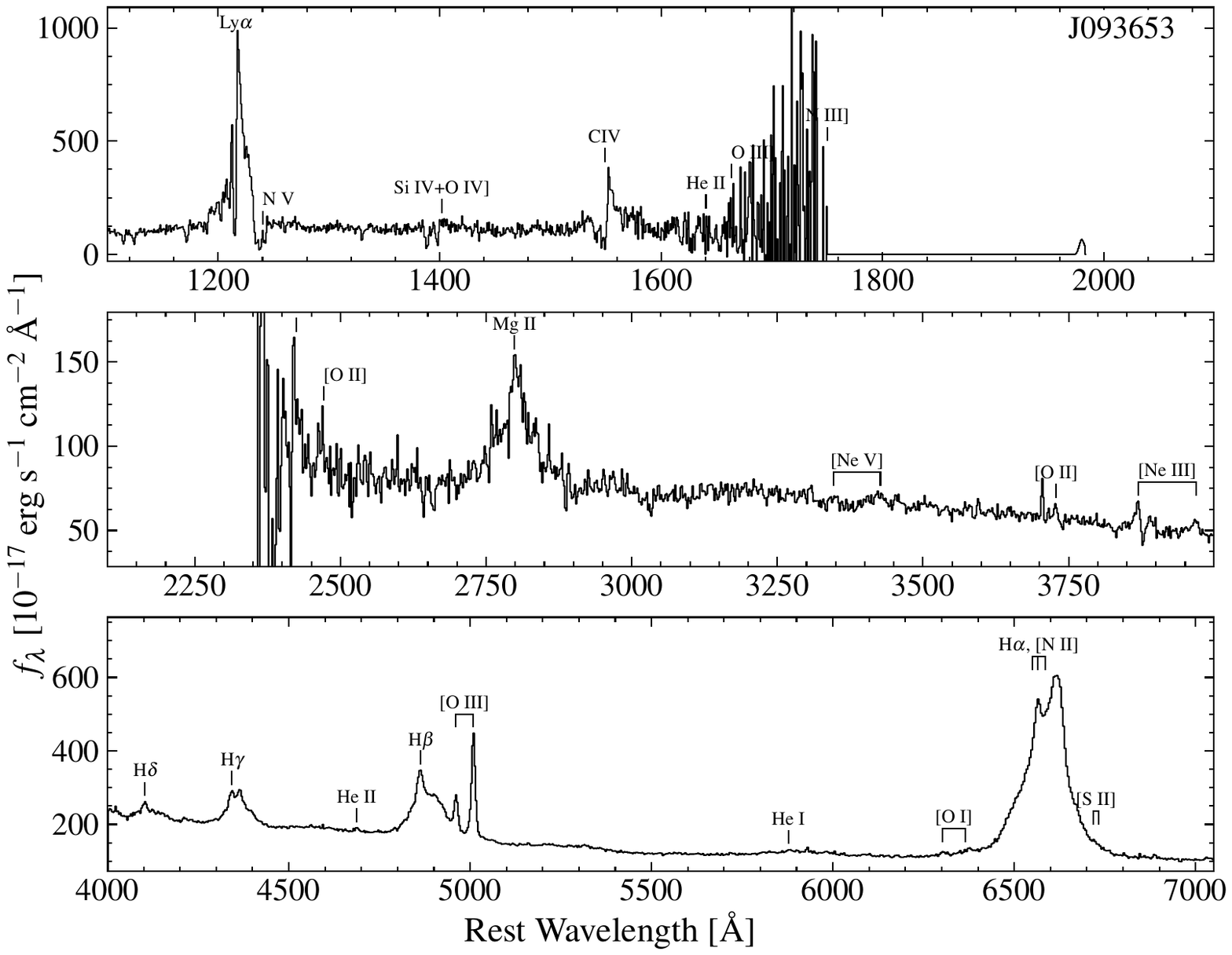} 	
}
\centerline{Figure~\ref{fig:fullspec}. -- Continued.}
\end{figure*}

\begin{figure*}
\centerline{
    \includegraphics[width=0.75\textwidth]{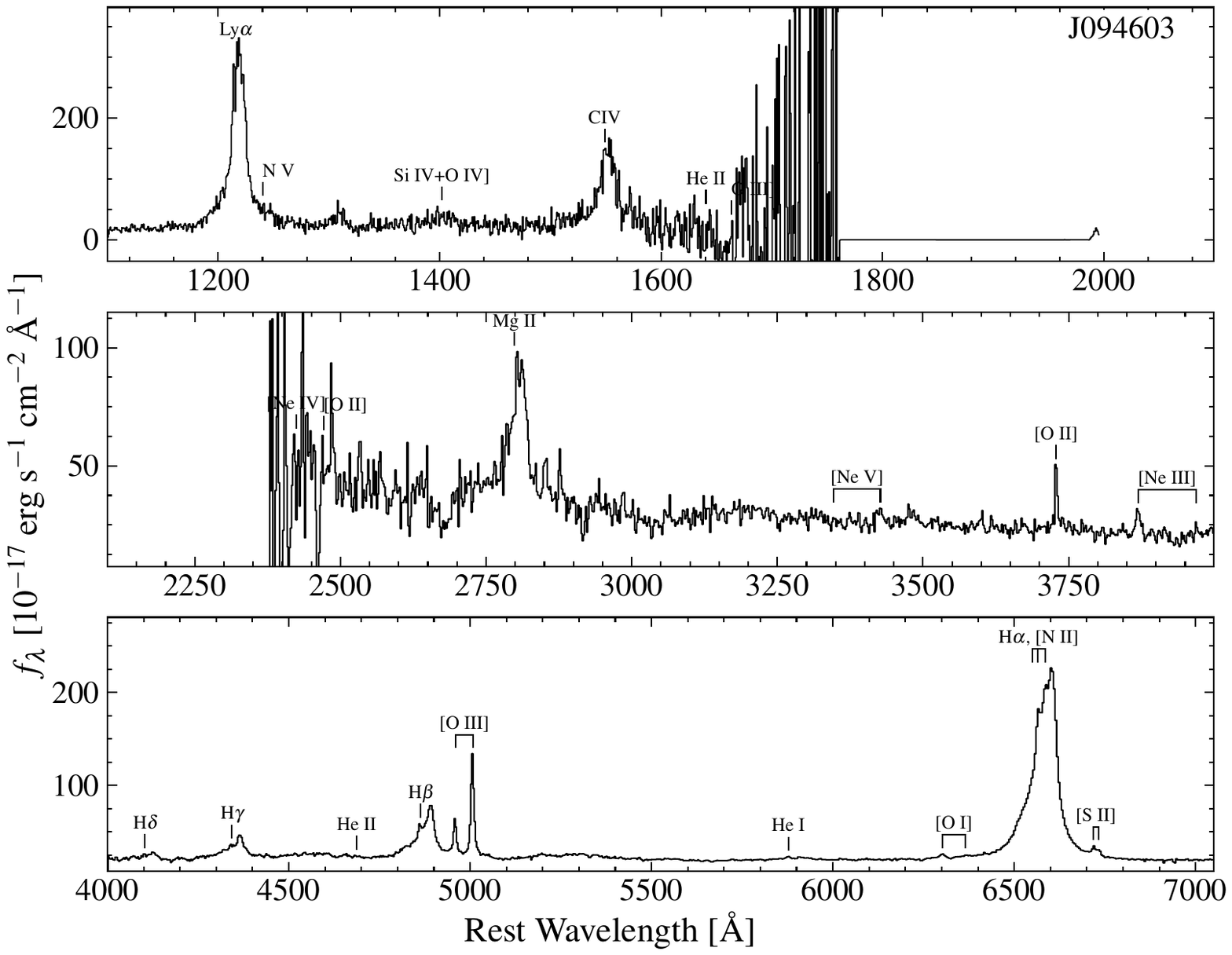} 	
}
\centerline{  
    \includegraphics[width=0.75\textwidth]{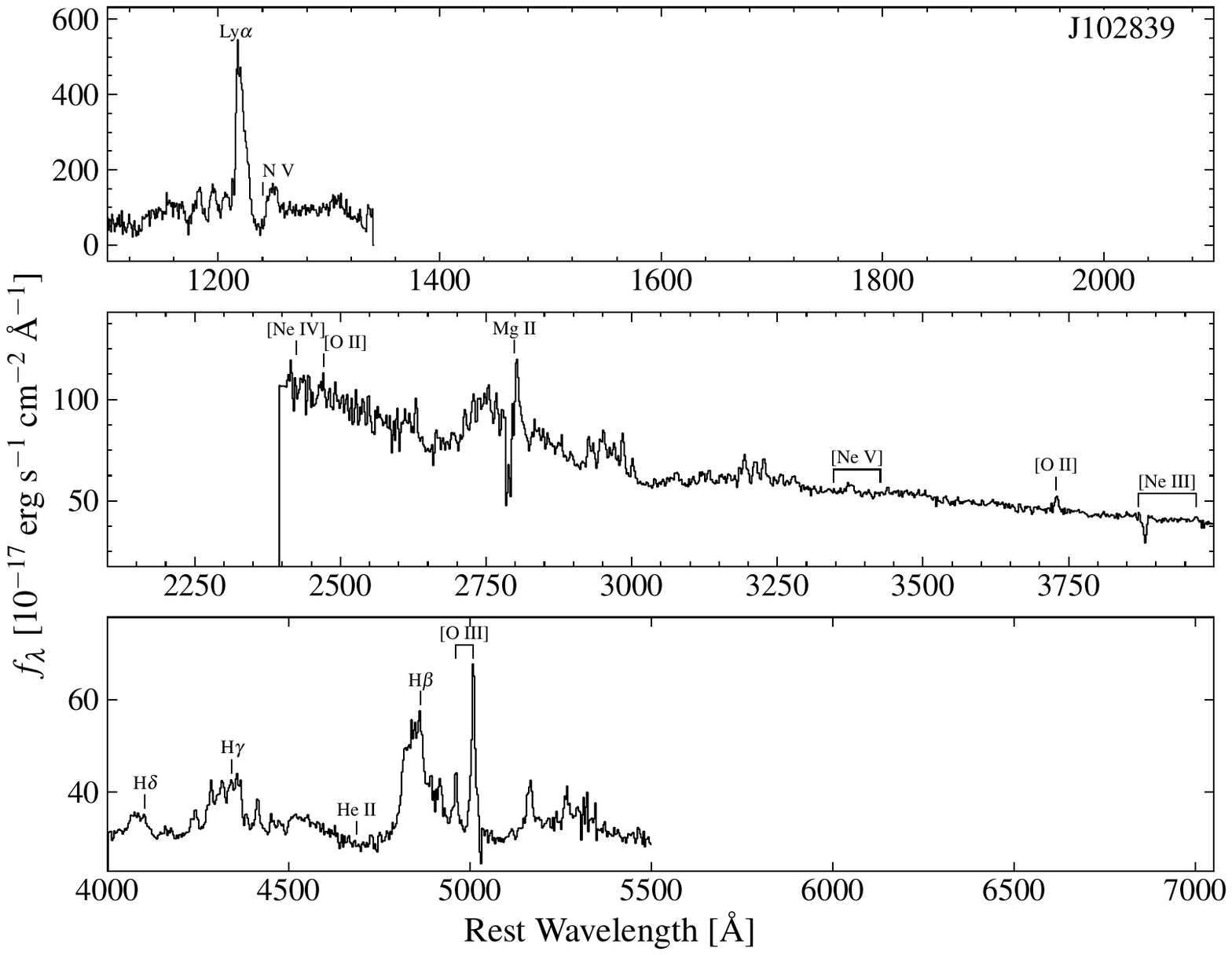} 	
}
\centerline{Figure~\ref{fig:fullspec}. -- Continued.}
\end{figure*}

\begin{figure*}
\centerline{
    \includegraphics[width=0.75\textwidth]{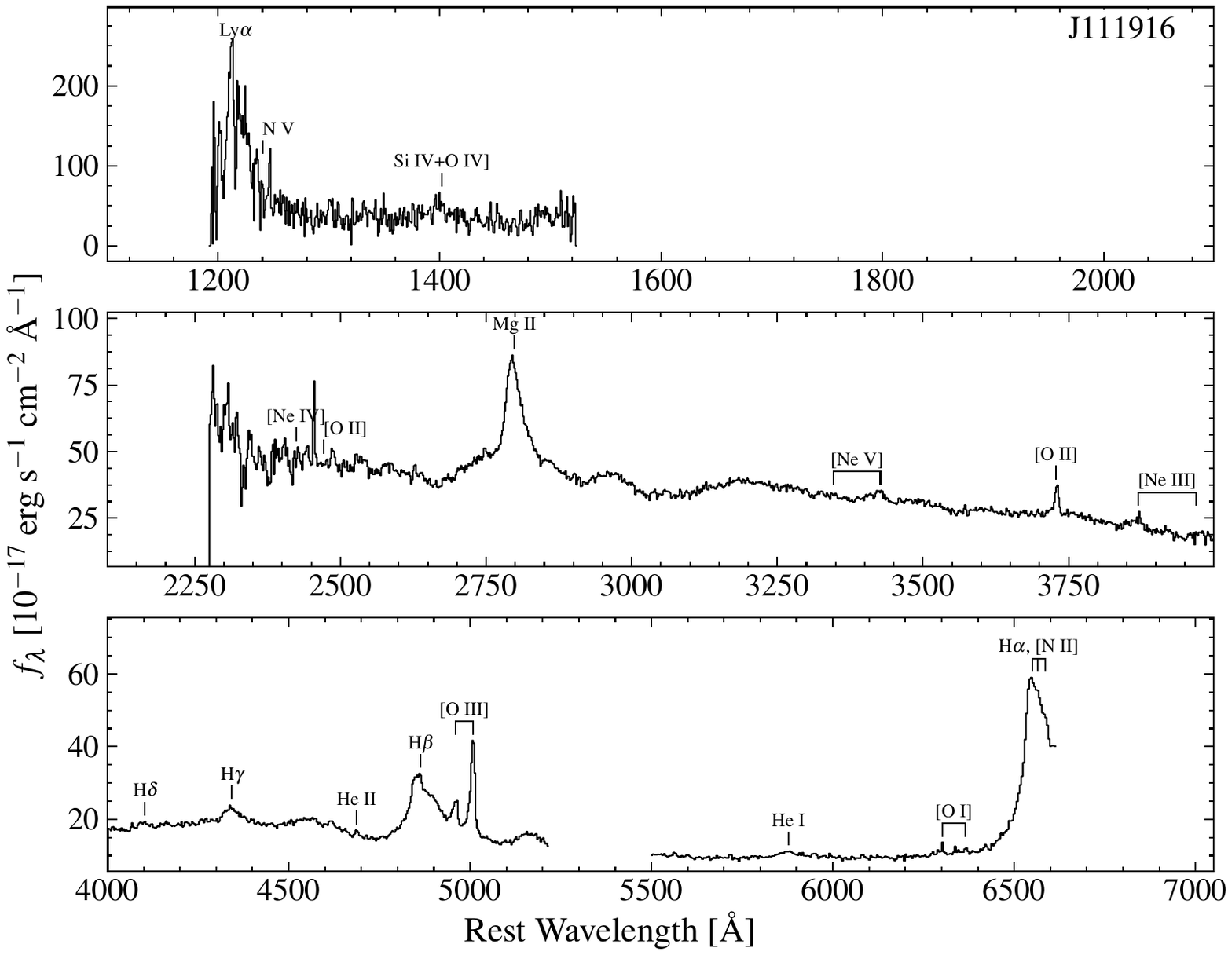} 	
}
\centerline{  
    \includegraphics[width=0.75\textwidth]{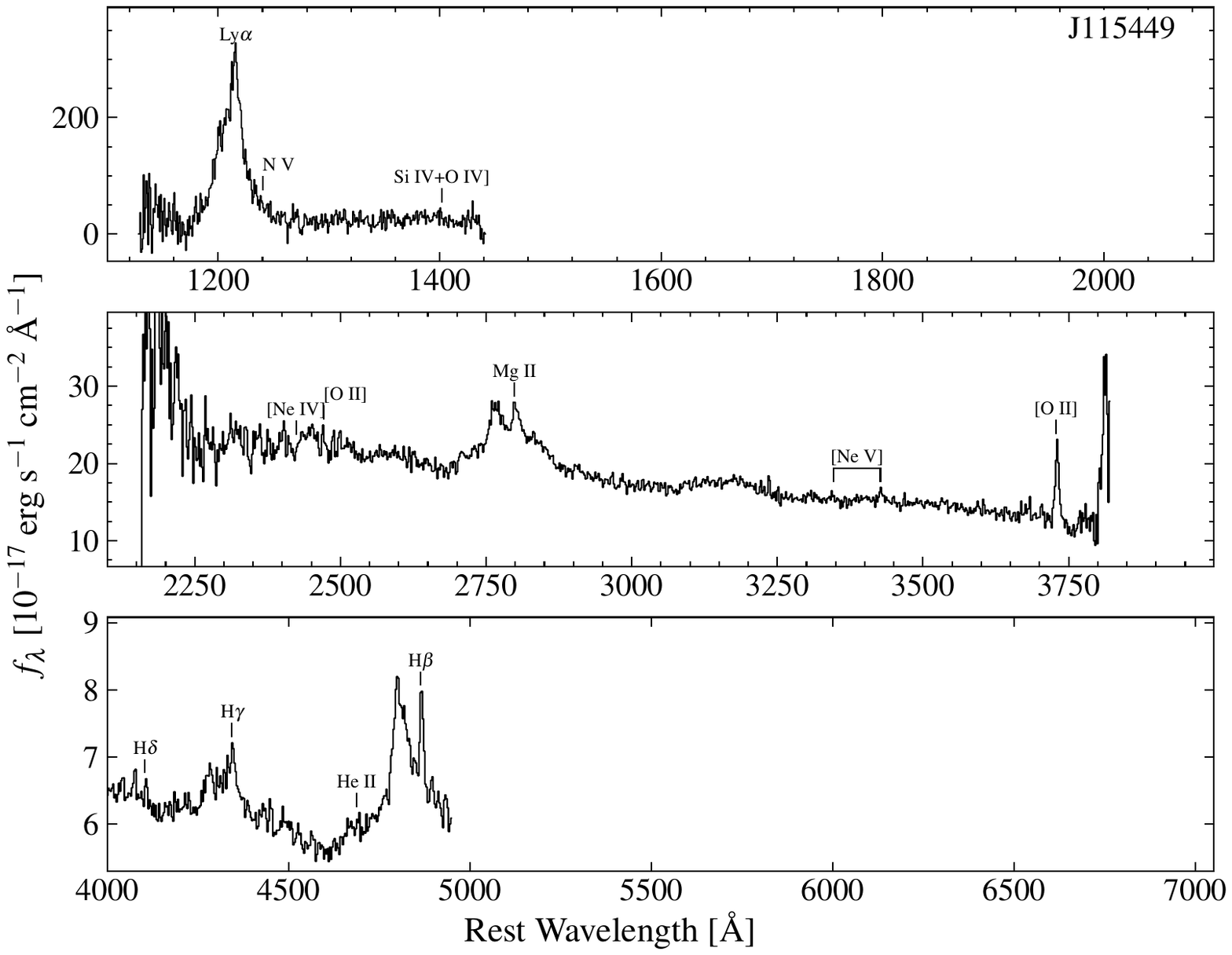} 	
}
\centerline{Figure~\ref{fig:fullspec}. -- Continued.}
\end{figure*}

\begin{figure*}
\centerline{
    \includegraphics[width=0.75\textwidth]{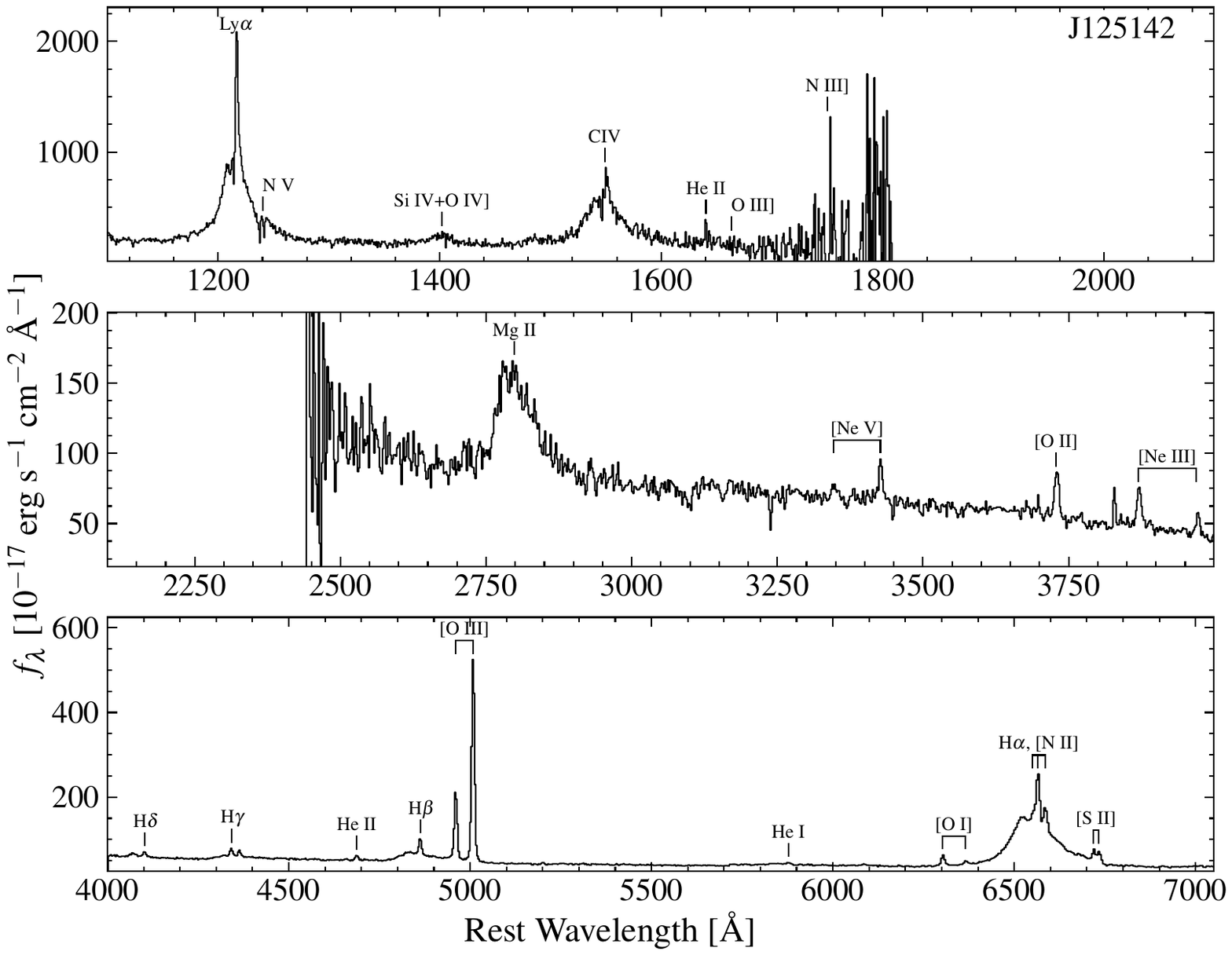} 	
}
\centerline{  
    \includegraphics[width=0.75\textwidth]{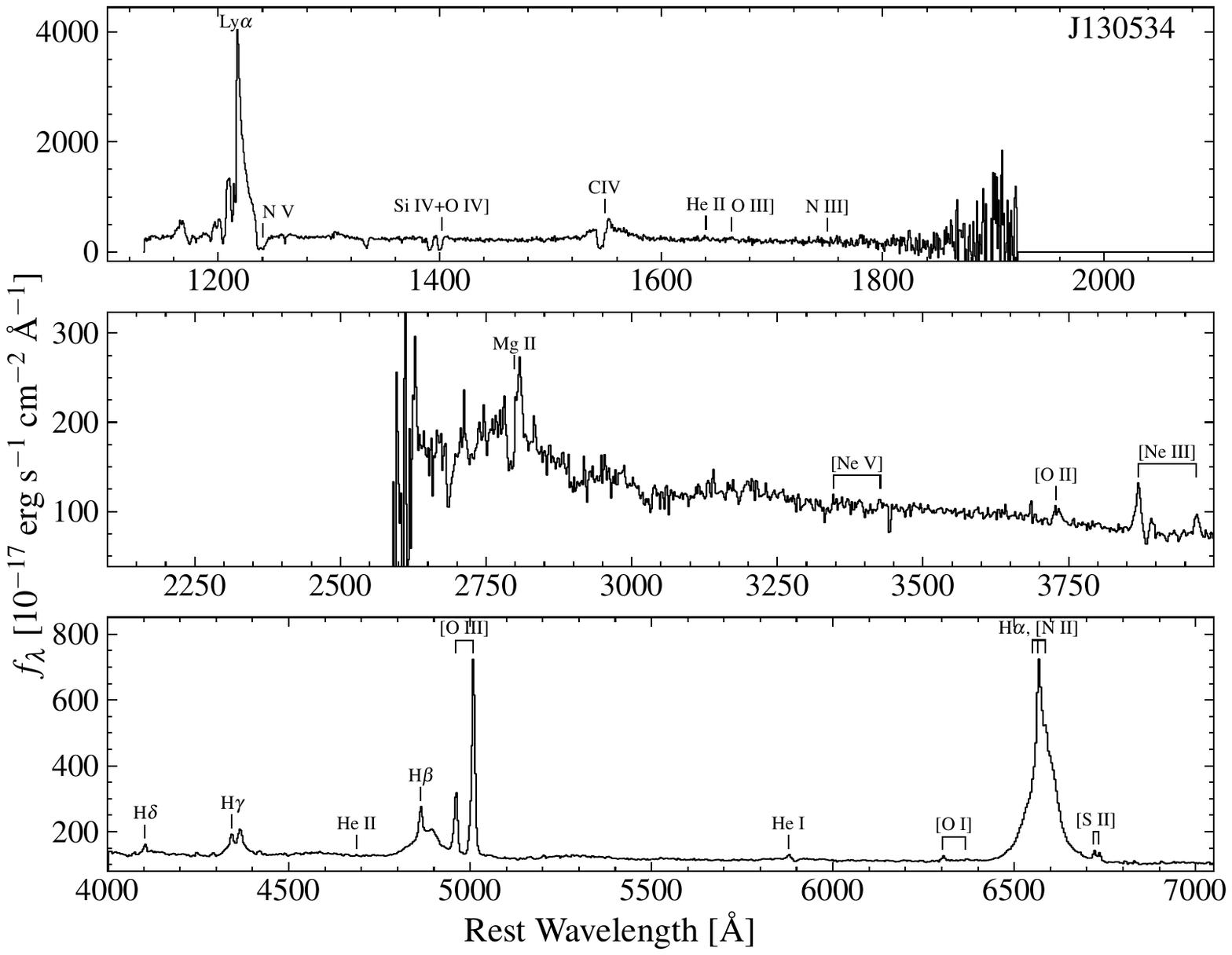} 	
}
\centerline{Figure~\ref{fig:fullspec}. -- Continued.}
\end{figure*}

\begin{figure*}
\centerline{
    \includegraphics[width=0.75\textwidth]{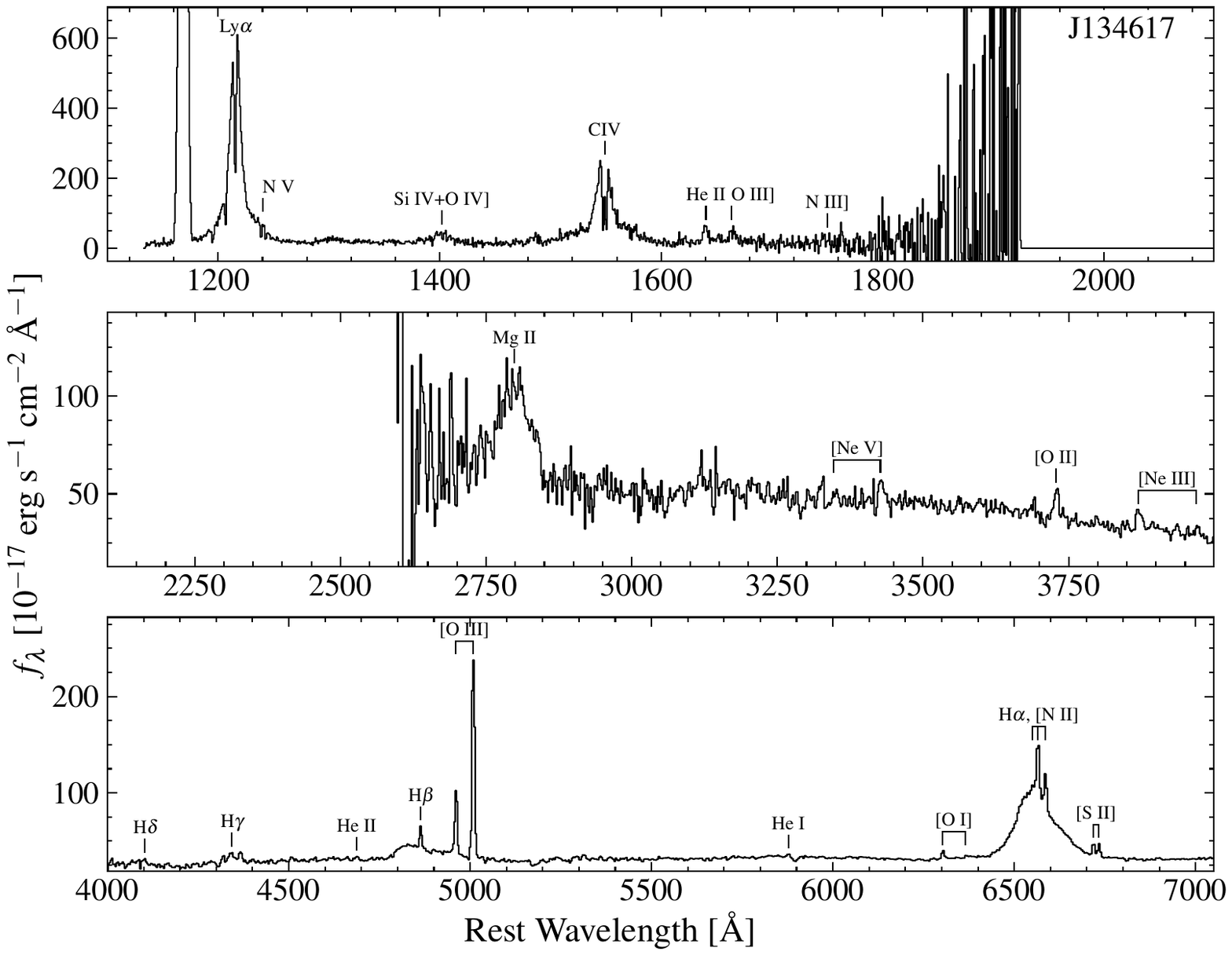} 	
}
\centerline{  
    \includegraphics[width=0.75\textwidth]{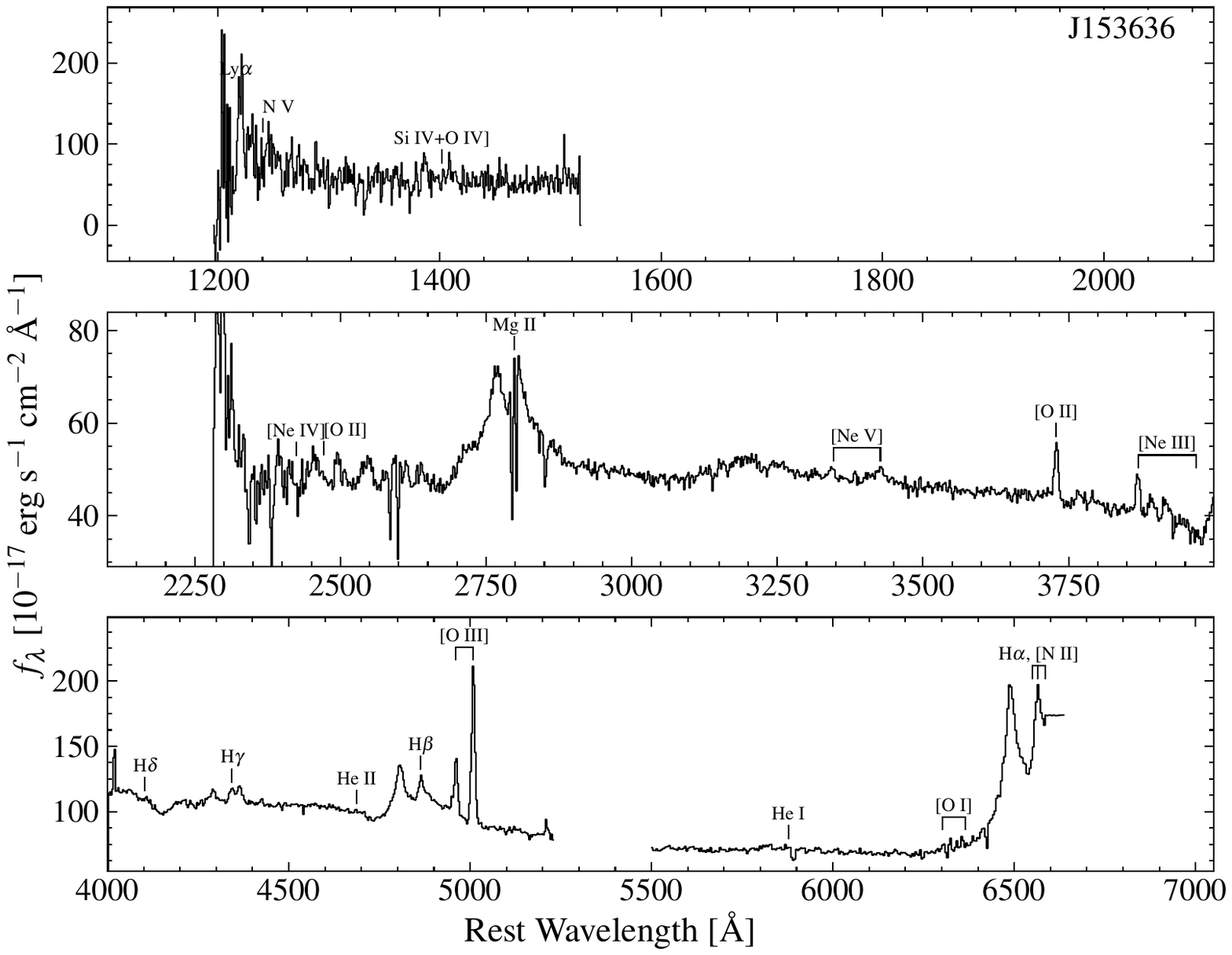} 	
}
\centerline{Figure~\ref{fig:fullspec}. -- Continued.}
\end{figure*}

\begin{figure*}
\centerline{
    \includegraphics[width=0.75\textwidth]{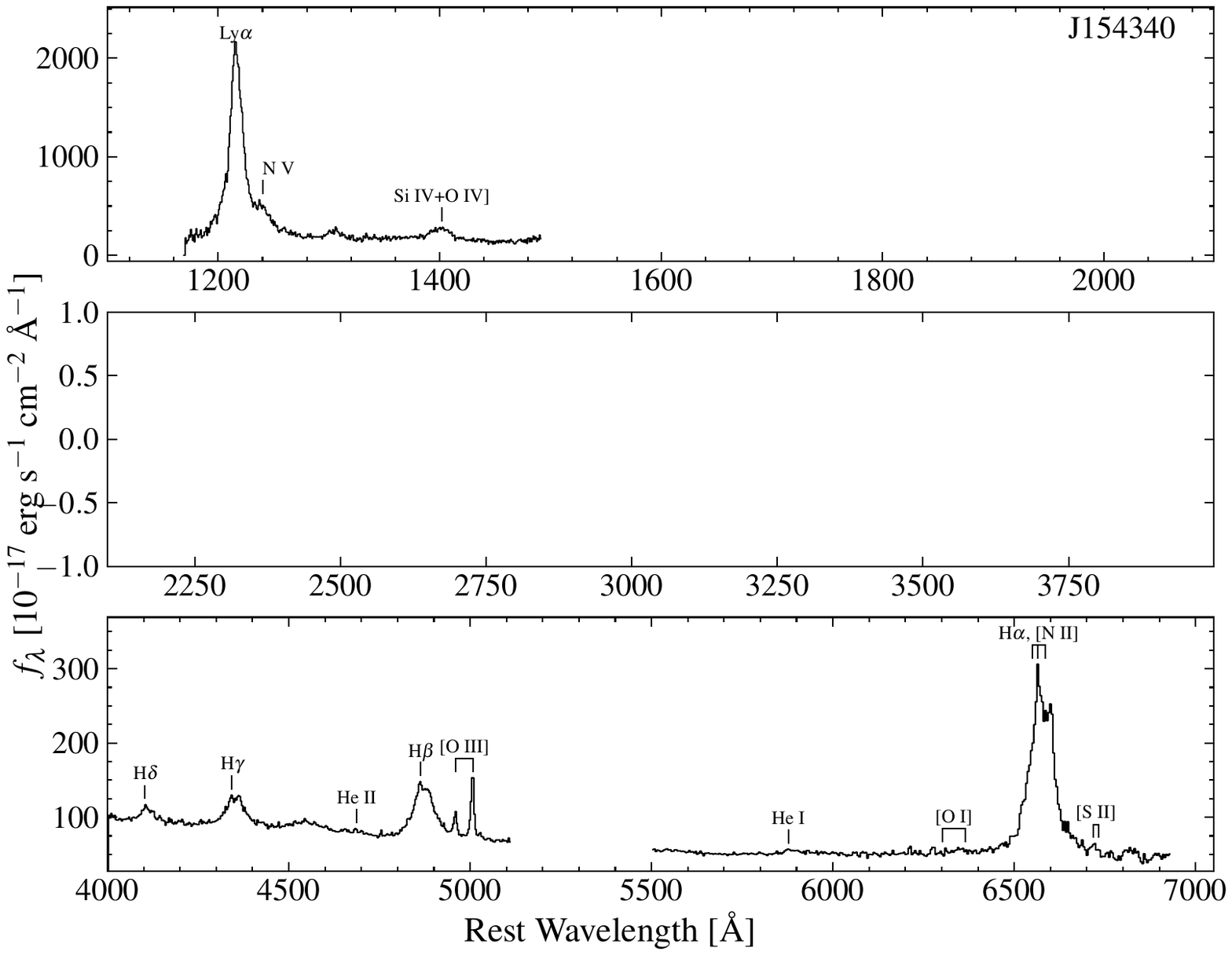} 	
}
\centerline{  
    \includegraphics[width=0.75\textwidth]{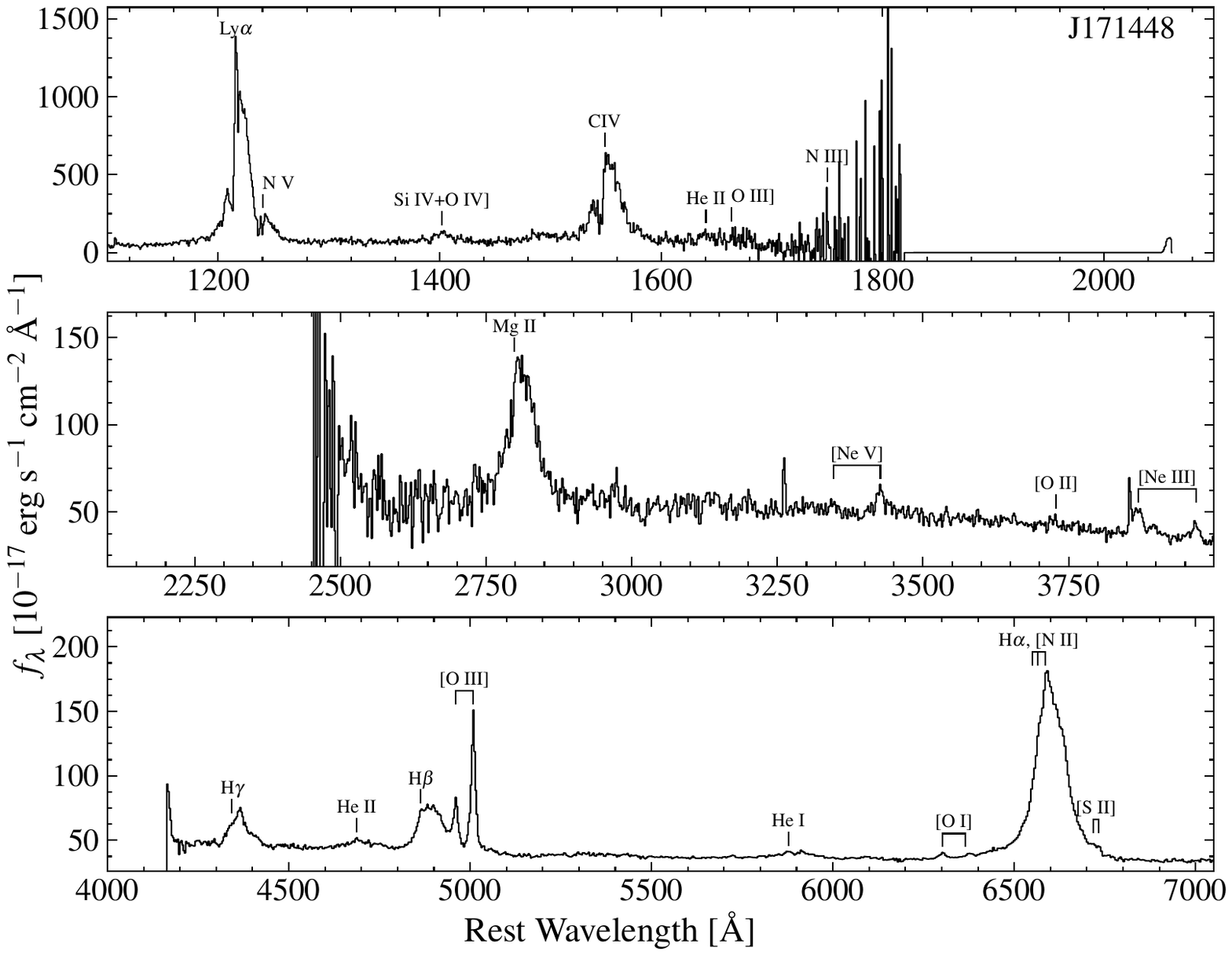} 	
}
\centerline{Figure~\ref{fig:fullspec}. -- Continued.}
\end{figure*}

\section{Analysis}
\label{sec:analysis}
The primary goal of this work is to compare the line profiles of the optical and UV profiles in candidate SBHBs. To facilitate this comparison, we performed a spectral decomposition to isolate the line profiles and derived scale factors that allowed us to normalize the profiles in order to compare their shapes and velocities.

\subsection{Spectral decomposition and measurements}
\label{sec:sfit}
In order to isolate the broad profiles of the \Lya, \CIV, \MgII, \Hb, and \Ha\ emission lines, we performed a spectral decomposition. We deblended each line of interest separately with \textsc{specfit} \citep{kriss94}, which fits parametric models to data by minimizing the $\chi^2$ statistic.

The UV emission lines are not blended with unambiguous narrow lines \citep{wills93b}, so we primarily needed to determine and subtract an appropriate model for the continuum in order to isolate the broad-line profile. For \Lya\ and \CIV\ we adopted a power law to model the continuum in line-free wavelength windows on either side of the broad lines. For \MgII, we modeled the continuum with a power law and UV \FeII\ template \citep{vestergaard01}, which was allowed to have zero flux if that provided the best fit. 

Isolating the broad \Hb\ profile is a more involved exercise, especially in this sample with unusual line profiles, because of contamination from optical \FeII, \OIIIdblt, and a narrow \Hb\ component. The spectral decomposition for this emission line was done in a three-stage process described in detail by \citet{runnoe15} and \citet{runnoe21}. The components of the model include a power law and \FeII\ template \citep{veron-cetty04} to model the continuum and a suite of Gaussians to parameterize the broad \Hb, narrow \Hb, and narrow \OIII\ emission lines. During the fit, we constrained the central wavelengths of the Gaussians used to parameterize the narrow emission lines to preserve the wavelength ratios between the lines.

Finally, we deblended the \Ha\ emission line. This model included a power law for the continuum, plus up to 4 Gaussians each to describe the \OIdblt, \NIIdblt, \SIIdblt, narrow \Ha, and broad \Ha\ line profiles (the narrow lines never required more than 2 Gaussians).

\begin{figure*}
\hspace{0.76cm}
\begin{minipage}{\columnwidth}
  \includegraphics[width=0.96\textwidth]{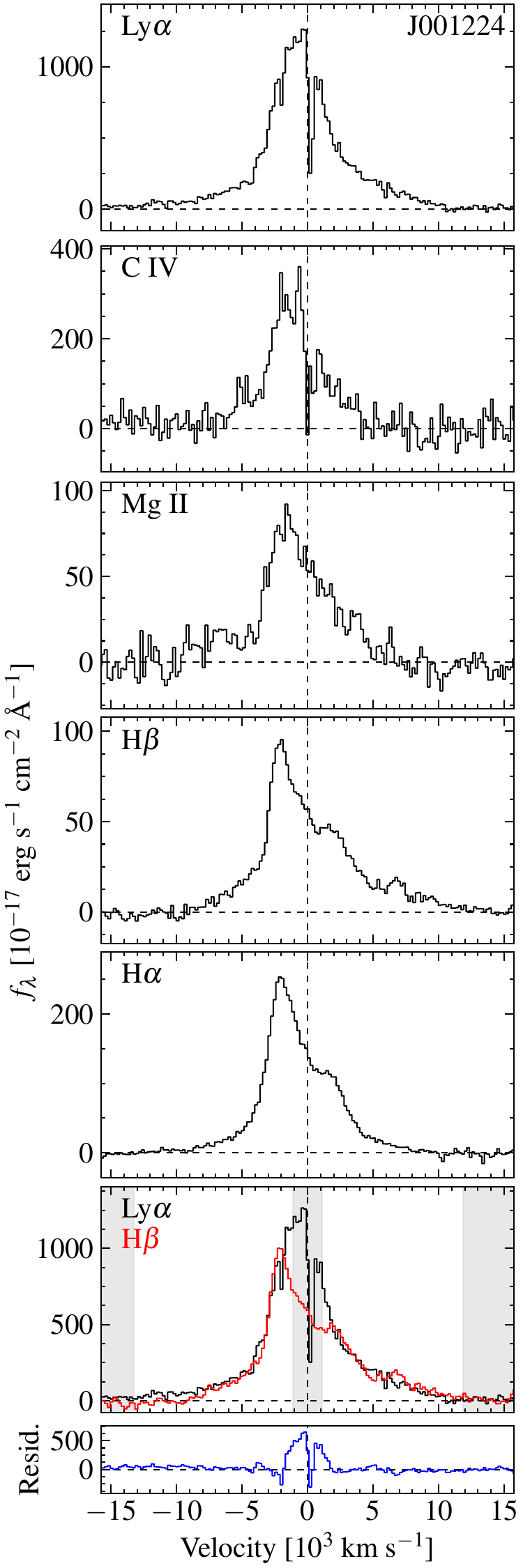} 
\end{minipage}
\begin{minipage}{\columnwidth}
\caption{
Isolated broad emission line profiles for the SBHB candidates. Dashed lines indicate where the flux density and velocity are zero, with zero velocity determined by the nominal wavelength of each emission line in the rest frame set by narrow [\ion{O}{3}]. The profile is derived in each case by subtracting the parametric model for the continuum and other narrow-line components from the spectrum, leaving only the broad line. Some of the line profiles have been binned to improve the S/N. The bottom two panels show a comparison of the continuum-subtracted \Hb\ (red) and Ly$\alpha$ (black) broad line profiles. The \Hb\ profile is scaled to match Ly$\alpha$ according to the methodology described in Section~\ref{sec:scale}. The gray regions were masked out when outlier rejection was performed to calculate the scale factor. J153636 was too absorbed to determine the scale factor via outlier rejection, so it was determined based on visual inspection. The \ion{N}{5}~$\lambda\lambda$1238,1242 emission lines are sometimes visible in the UV spectra at  6,000~\kms, respectively. The geocoronal \Lya\ in J13417 is also evident at $-12,000$~\kms.  The broad \ion{He}{2} line is sometimes visible in the optical spectrum around $-11,000$~\kms. The residual (where the scaled \Hb\ is subtracted from Ly$\alpha$) is shown in blue. }
\label{fig:vstack}
\end{minipage}
\end{figure*}

In Figure~\ref{fig:vstack} we present the broad line profiles for objects observed with the HST. For each line, the continuum (consisting of the power-law model and, when appropriate, the \FeII\ templates) has been subtracted, as well as any optical narrow emission lines. The procedure for subtracting components from the UV line profiles is less clear. While the optical lines have distinct broad and narrow emission lines that can be ascribed to separate physical regions \citep[e.g.,][]{shang05}, the UV line profile cores typically have an intermediate width core of disputed origin. 
It has been ascribed to an intermediate line region (ILR) which sits spatially between the BLR and narrow-line region \citep{wills93b,brotherton94} or a disk wind \citep{richards11}. We opt for a conservative approach and do not subtract any component of the UV line profiles. 

\begin{figure*}
\centerline{
  \includegraphics[width=0.45\textwidth]{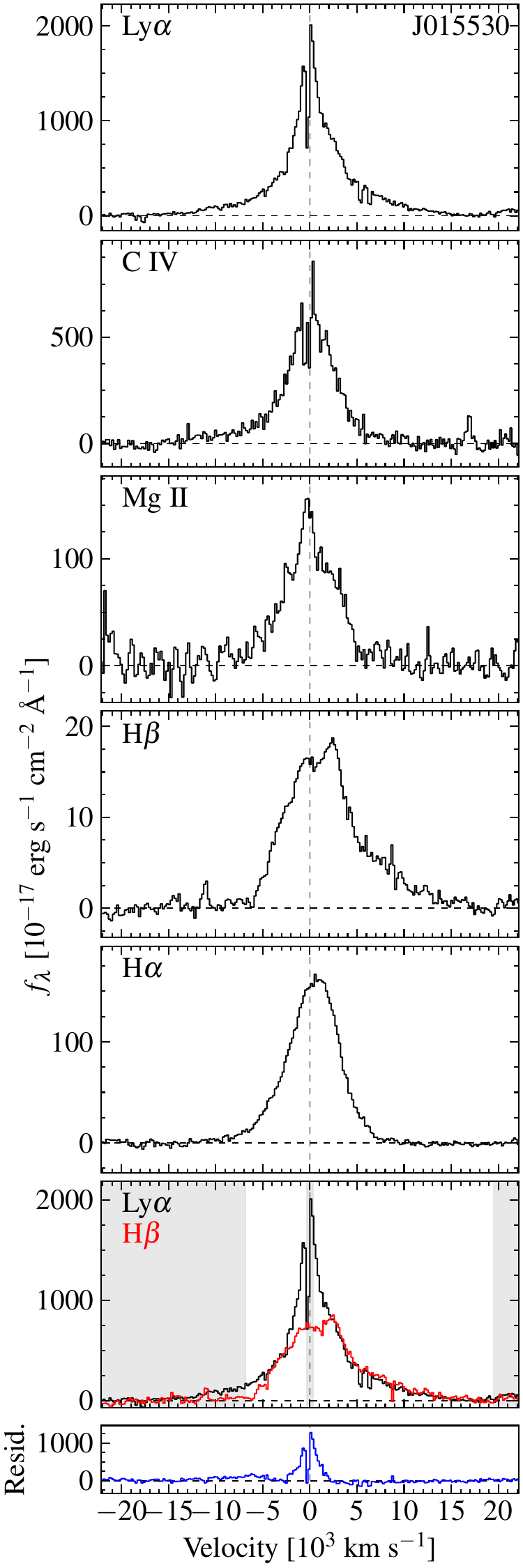} 
  \includegraphics[width=0.45\textwidth]{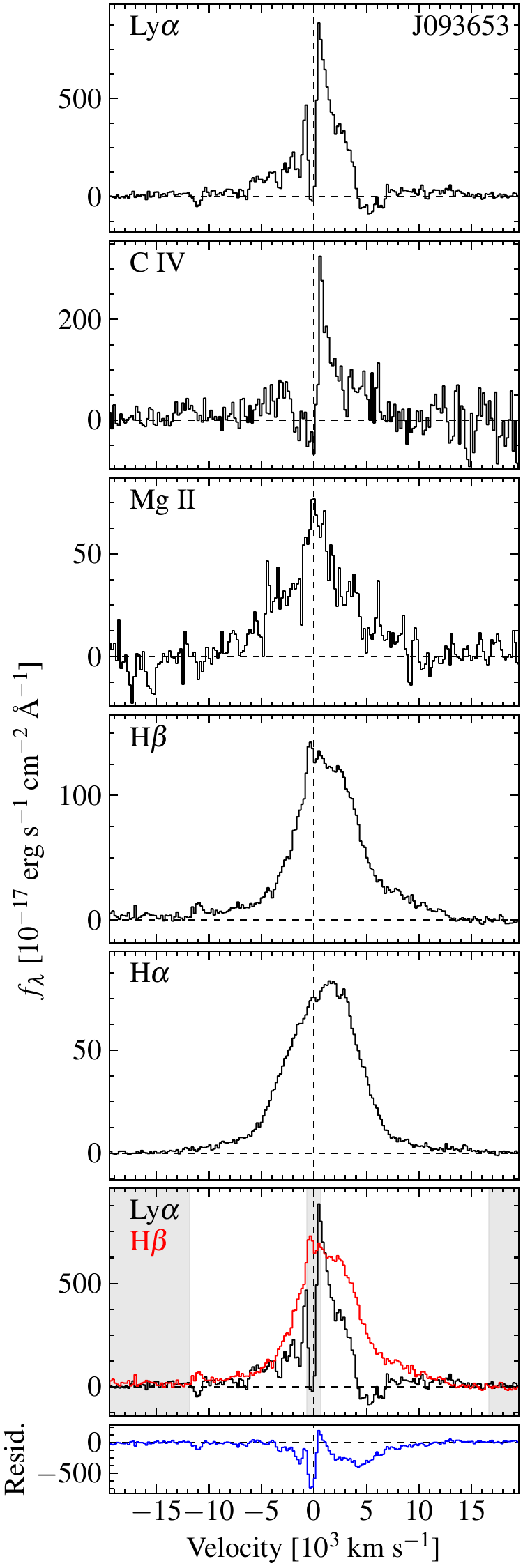} 	
}
\centerline{Figure~\ref{fig:vstack}. -- Continued.}
\end{figure*}
\begin{figure*}
\centerline{
  \includegraphics[width=0.45\textwidth]{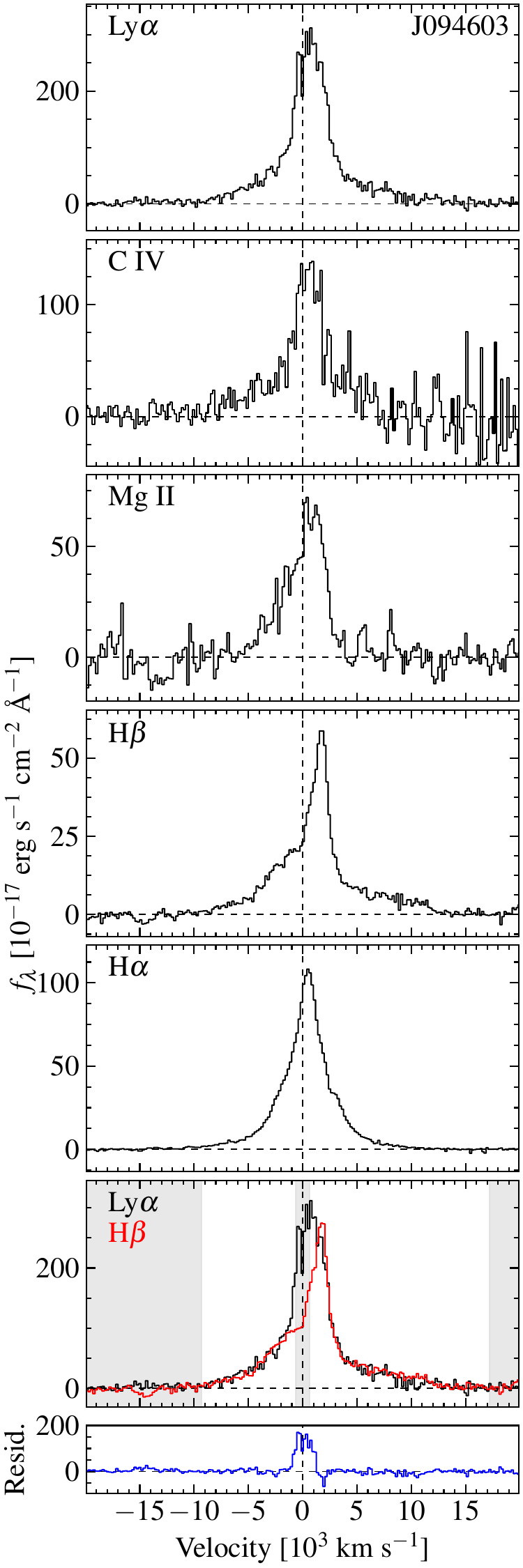} 	
  \includegraphics[width=0.45\textwidth]{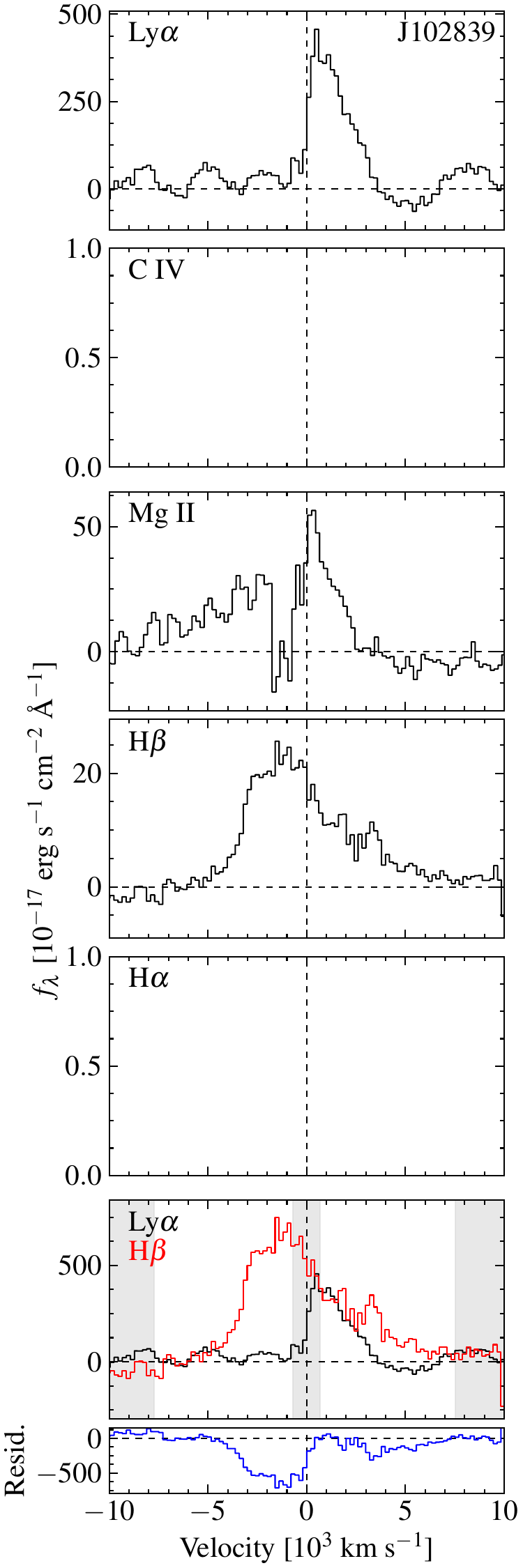} 	
}
\centerline{Figure~\ref{fig:vstack}. -- Continued.}
\end{figure*}

\begin{figure*}[!t]
\centerline{
  \includegraphics[width=0.45\textwidth]{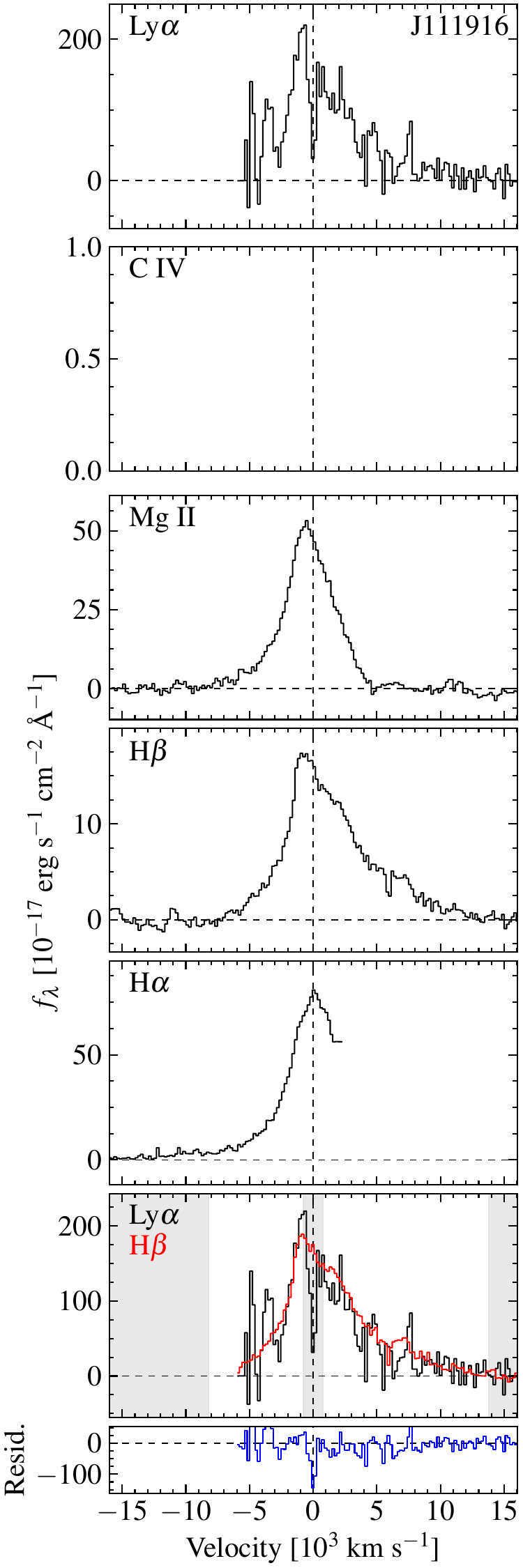} 	
  \includegraphics[width=0.45\textwidth]{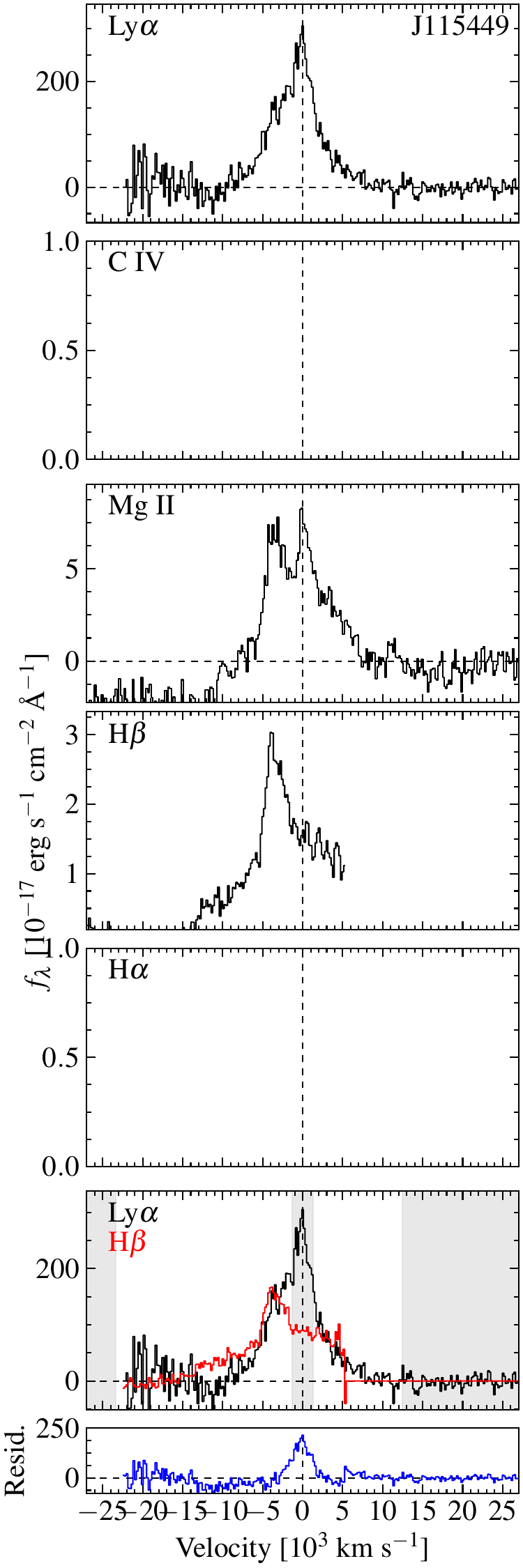} 	
}
\centerline{Figure~\ref{fig:vstack}. -- Continued.}
\end{figure*}
\begin{figure*}
\centerline{
  \includegraphics[width=0.45\textwidth]{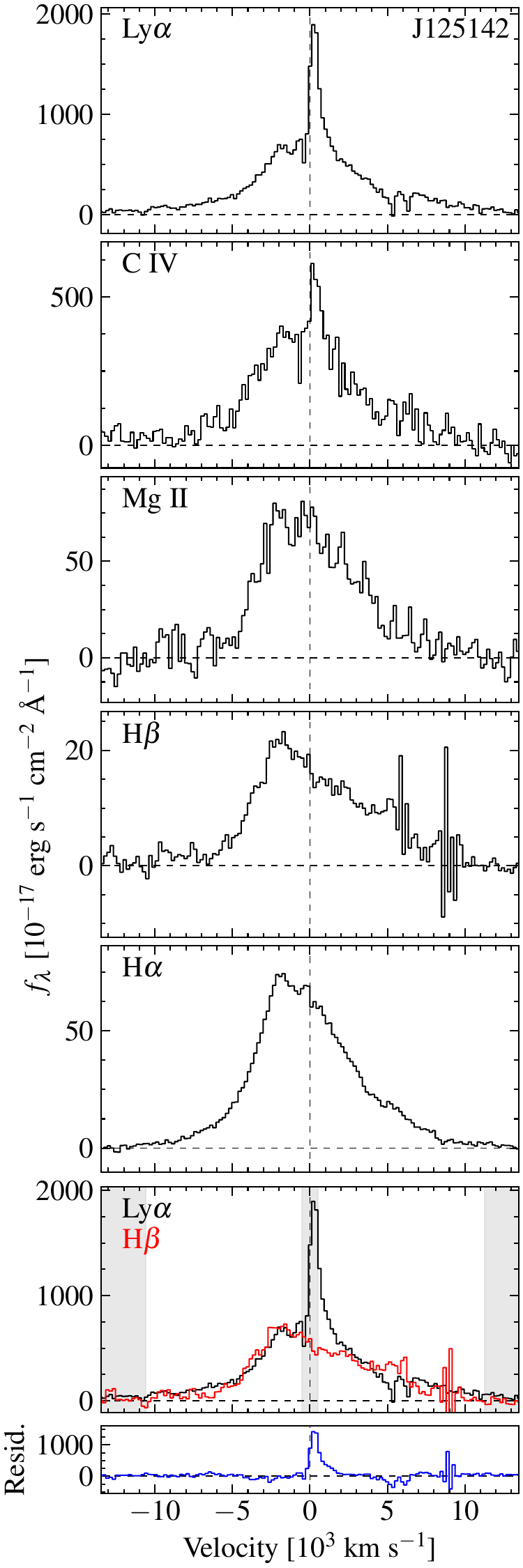} 	
  \includegraphics[width=0.45\textwidth]{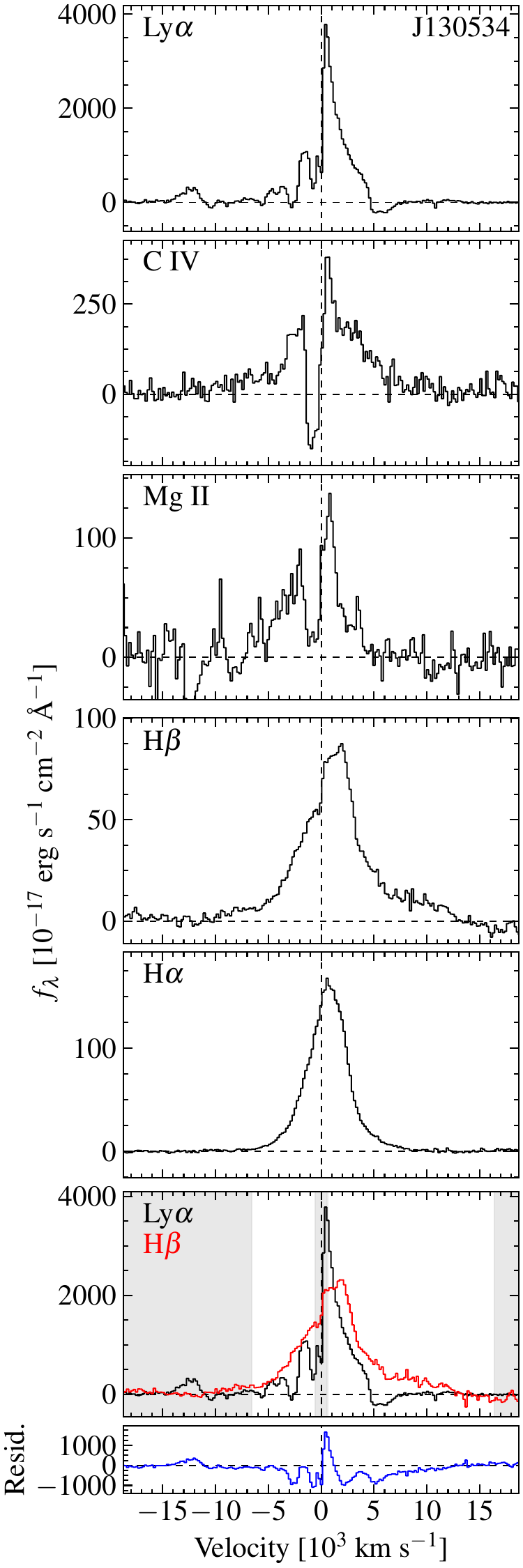} 	
}
\centerline{Figure~\ref{fig:vstack}. -- Continued.}
\end{figure*}
\begin{figure*}
\centerline{
  \includegraphics[width=0.45\textwidth]{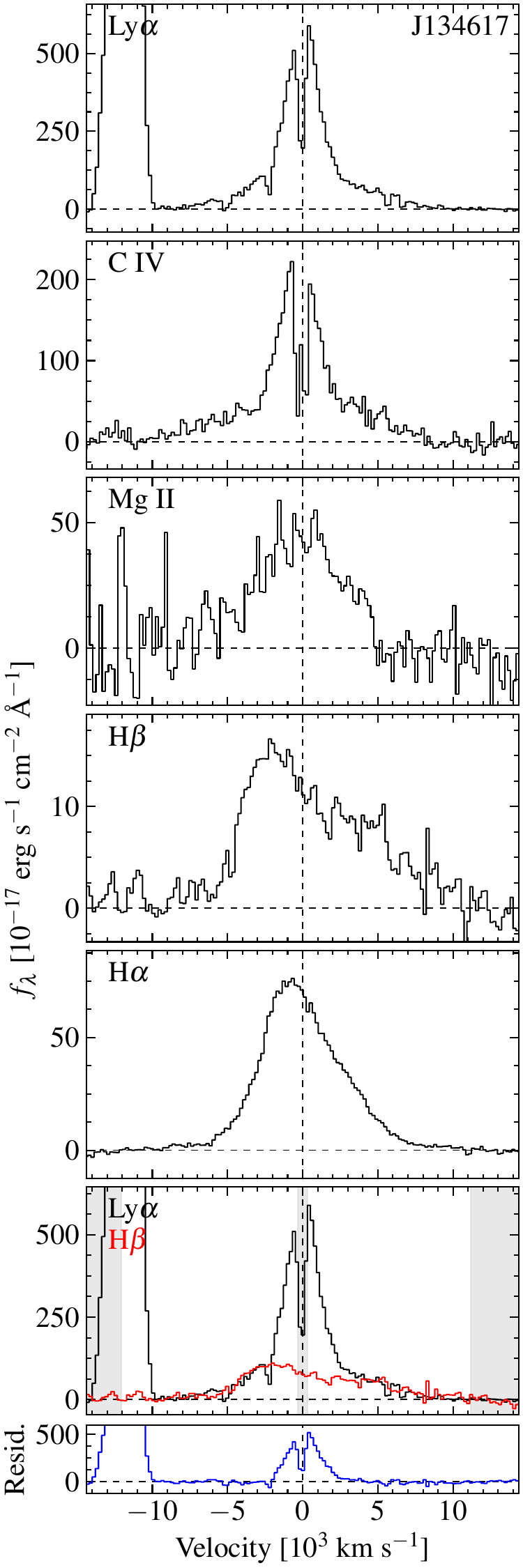} 	
  \includegraphics[width=0.45\textwidth]{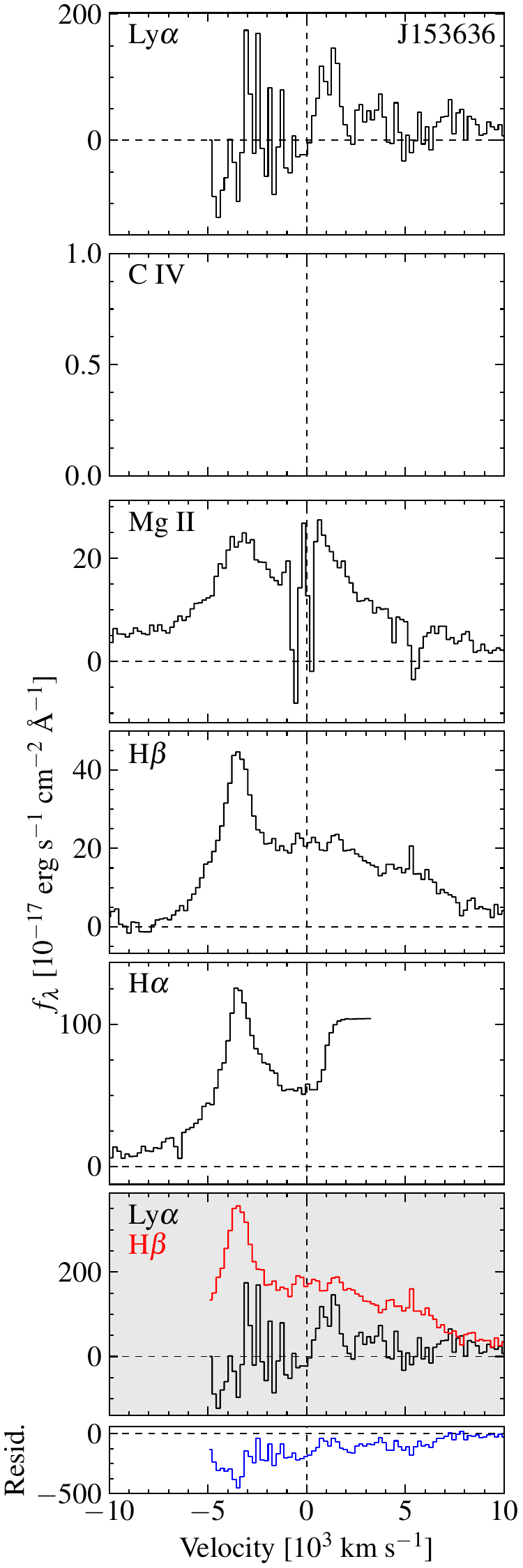} 	
}
\centerline{Figure~\ref{fig:vstack}. -- Continued.}
\end{figure*}
\begin{figure*}
\centerline{
  \includegraphics[width=0.45\textwidth]{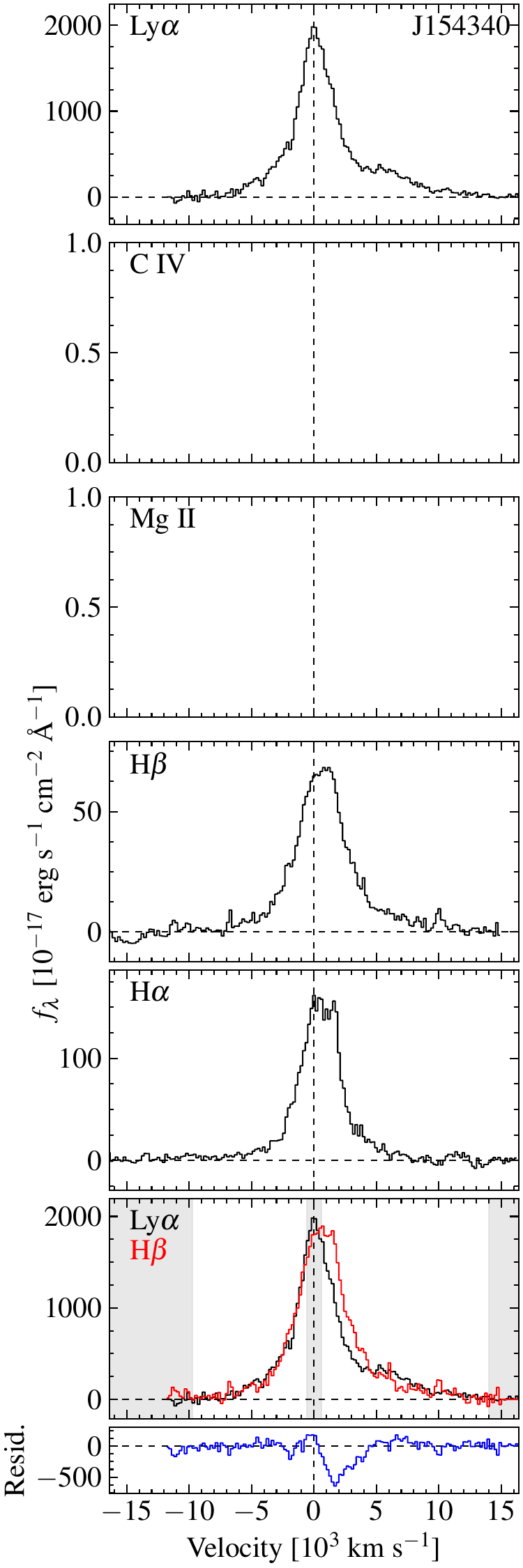} 	
  \includegraphics[width=0.45\textwidth]{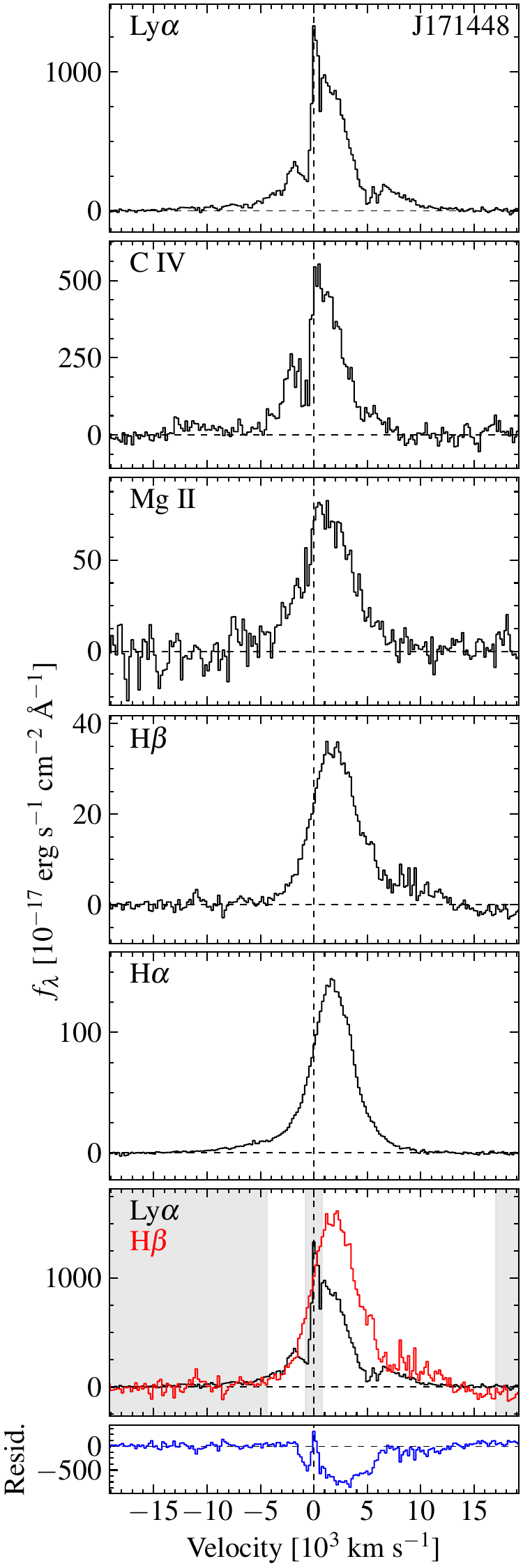} 	
}
\centerline{Figure~\ref{fig:vstack}. -- Continued.}
\end{figure*}

\subsection{Scale factors for matching line profiles}
\label{sec:scale}
In order to compare the optical and UV broad line profiles further, we determined a scale factor to match them. This was done via the following procedure. First, we identified the wavelength region over which the line profiles should be compared. This was taken to be the wavelength window where the broad \Hb\ emission line was above 1\% of the peak value determined after the spectral decomposition. This is intended to be inclusive; even though fluxes at 1\% of the broad-line peak may be lost in the noise, as much of the line profile as possible is being adopted for a comparison. 

Next, we masked a region around the \Hb\ narrow line. This window was determined by the wavelengths where the \Hb\ narrow component from the spectral decomposition was at least 10\% of its peak value, leaving as much as possible for broad-line comparison. 

Finally, we determined the best multiplicative scale factor via a least squares approach with iterative sigma clipping. The sigma clipping algorithm was included to reject points which need not be well described by a scale factor, in this case typically corresponding to noisy pixels, pixels in the core of the UV line profile, or pixels impacted by absorption in the UV line profile. The scaled \Lya\ and \Hb\ profiles are shown together in the bottom of Figure~\ref{fig:vstack}. Here we also show the residual spectrum, which is calculated by subtracting the scaled \Hb\ profile from \Lya. The residual spectrum can be used to visually estimate any contribution from the ILR, which will appear as a velocity centered core in the residual spectrum.



\subsection{Comparing the line profiles}
\label{sec:comp}
We used the stacked and scaled line profiles in Figures~\ref{fig:vstack} to test the nature of the SBHB candidates. According to the test, for a single source of all broad emission lines to be associated with one black hole in an SBHB that undergoes bulk orbital motion, all of the lines must be offset by a similar amount and the Ly$\alpha$ and \ion{C}{4} lines should have similar profiles and offsets as \Hb. If the offset \Hb\ is one of the two peaks of a double-peaked line caused by a non-axisymmetric structure in the BLR around a single supermassive black hole, then by analogy with well studied DEs the corresponding Ly$\alpha$ and \ion{C}{4} lines can be single-peaked and centered on or near the velocity frame of the narrow emission lines.

Below we evaluate the credentials of each quasar and render a judgment that falls in one of the following categories:

\begin{itemize}
    \item[] {\bf Strong support:} The broad line profiles of the optical and UV lines all show consistent, velocity offset peaks in strong support of the SBHB hypothesis.
    \item[] {\bf Tentative support:} The broad line profiles of the optical and UV lines appear consistent, albeit with increased uncertainty due to a strong ILR component or weak absorption. This is tentative support for the SBHB hypothesis.
    \item[] {\bf Disfavored:} The profiles of the UV broad lines do not match the velocity offset seen in the optical lines and the SBHB hypothesis is disfavored by this test. 
    \item[] {\bf Absorbed:} It is impossible to compare the optical and UV line profiles because the UV emission-line profiles are severely distorted by strong associated absorption lines (from \Lya\ or \ion{N}{5}).
\end{itemize}

We find evidence of all four scenarios in the data, which we summarize in Table~\ref{tab:class}. Four objects show strong support for the SBHB hypothesis, four show tentative support complicated by absorption and uncertainty about the ILR in single AGN, the SBHB hypothesis is disfavored in one object, and in four objects the absorption is so severe that it was not possible to make a meaningful comparison of the line profiles.

\begin{deluxetable}{cc}
\setlength{\tabcolsep}{4pt}
\tablecolumns{9}
\tablewidth{\textwidth}
\tablecaption{UV support for SBHB hypothesis \label{tab:class}
}
\tablehead{
\colhead{} & 
\colhead{Primary} \\
\colhead{Object} &
\colhead{Classification\tablenotemark{a}} 
}
\startdata
J001224   	&    strong	    \\
J015530   	&    tentative  \\
J093653   	&    absorbed	\\
J094603   	&    tentative	\\
J102839	    &    absorbed \\
J111916    	&  	 tentative  \\
J115449	    &  	 strong    \\
J125142   	&    strong	    \\
J130534   	&    absorbed \\
J134617   	&    tentative	\\
J153636   	&    absorbed	\\
J154340   	&    disfavor	\\
J171448   	&    tentative	\\
\enddata
\tablenotetext{a}{Details in Section~\ref{sec:comp}.}
\end{deluxetable}

We note that some classifications depend on the interpretation of the ILR, because the line core must either be included in the comparison (if it originates in the BLR and therefore should undergo bulk orbital motion) or not (if it is emitted external to the BLR). In our classifications, we adopt the interpretation that the ILR line core is not emitted from the BLR and therefore will not exhibit signs of orbital motion. The primary motivation for this interpretation is the results of reverberation mapping campaigns, which show that the core of UV line profiles do not reverberate in response to changes in the continuum emission \citep{denney12}, indicating that it was not emitted from the same region that emits the broad optical lines and the base of the broad UV lines. Therefore, we allow a symmetric, intermediate-width core component to the UV line profiles to be excluded from our comparisons. In other words, optical and UV profiles are considered to match even if the UV profiles have a residual, so long as the residual is symmetric, intermediate in width (i.e. narrower than the broad lines themselves) and centered at zero velocity. Evidence of this component will appear in the residual panels of Figure~\ref{fig:vstack}.

\subsection{Notes on individual objects}
\label{sec:iobj}

\noindent \emph{J001224.--}
There is a very compelling blueshift of comparable magnitude is observed in all five line profiles and the \ion{Mg}{2} profile in particular is a very good match to \Hb. A comparison of \Lya\ and \Hb\ implies a symmetric ILR component in \Lya. This is a case of strong support for the SBHB hypothesis.

\noindent \emph{J015530.--}
The UV line profiles have a strong ILR component, but the broad base of the UV line profiles share a redshifted peak with the optical broad lines. This is a case where the SBHB hypothesis is tentatively supported under the ILR description of the UV lines.

\noindent \emph{J093653.--}
The Ly$\alpha$, \ion{C}{4}, and \Hb\ lines are observed in this candidate. Scaling the blue wing of the broad \Hb\ line to Ly$\alpha$ implies that the entire red wing of the Ly$\alpha$ profile, including its peak, has been absorbed. This is a case where the absorption is too strong to draw meaningful comparisons between the line profiles.

\noindent \emph{J094603.--}
The Balmer lines have very cuspy profiles that are not obviosuly as distinct in the UV line profiles, especially \ion{C}{4} and Ly$\alpha$. However, all the line profiles are redshifted to some degree and the subtraction of the scaled \Hb\ spectrum from \Lya\ leaves an ILR-like residual indicating the shape difference in cuspiness may be a red herring. This is a case of tentative support for the SBHB hypothesis.

\noindent \emph{J102839.--}
Ly$\alpha$ is absorbed over much of the line profile, including where the broad \Hb\ peaks, and the \ion{N}{5} and \ion{Mg}{2} profiles show absorption at the same velocities. This is a case where the absorption is too severe to compare the line profiles.

\noindent \emph{J111916.--}
The Ly$\alpha$ and H$\alpha$ profiles do not have full coverage, but the line peak is resolved in all profiles. The UV line profiles are noisy and have several narrow absorption lines that introduce ambiguity, but all of their shapes and peaks appear consistent with \Hb. The H$\alpha$ profile does not appear to match \Hb\ very well; in the raw spectrum the shift appears clear so this may be a case where the narrow-line removal is not optimal given the incomplete wavelength coverage. This is a case of tentative support for the SBHB hypothesis.

\noindent \emph{J115449.--}
The H$\beta$ profile is only partially covered, but the blueshifted peak is resolved. The \ion{Mg}{2} profile matches \Hb\ very well. The Ly$\alpha$ profile is noisy, and matches \Hb\ under the ILR description of the UV lines. In that case, the hump on the blue shoulder of the line is the blue peak that matches \Hb. Here, the shapes are slightly different, but the blueshifts are shared. This is a case of strong support for the SBHB hypothesis.

\noindent \emph{J125142.--}
The \Hb\ profile has a distinct, blueshifted peak that is very well matched in every other broad line profile, modulo an ILR core in the Ly$\alpha$ and \ion{C}{4} profiles marked by clear inflection points. This is a case of strong support for the SBHB hypothesis.

\noindent \emph{J130534.--}
The \Hb\ and \Ha\ profiles show distinct, redshifted peaks. All the UV line profiles, including \ion{N}{5} which is on the red wing of Ly$\alpha$, suffer absorption with multiple blueshifted troughs. This is a case where the absorption is too severe for a meaningful comparison of the line profiles.


\noindent \emph{J134617.--}
The \ion{C}{4} and \Lya\ line profiles have strong ILR cores that complicate the comparison to the optical line profiles. However, the bases of these lines, and also the \ion{Mg}{2} line all appear to share a blueshift. This is a case of tentative support for the SBHB hypothesis.

\noindent \emph{J153636.--}
The H$\alpha$ line has only partial coverage, but includes the very cuspy blueshifted peak. The \ion{Mg}{2} line profile matches \Hb\ well. The Ly$\alpha$ line is so absorbed, it is not possible to compare it to the optical line profiles. This is a case where \ion{C}{4} is not observed and Ly$\alpha$ is so absorbed that there is no clear conclusion.

\noindent \emph{J154340.--}
When \Hb\ is scaled to the blue wing of Ly$\alpha$, it implies absorption over the entire red wing of the line. There is no sign of absorption in Ly$\alpha$ and in fact, the \ion{N}{5} line is seen clearly in emission and the profile is otherwise smooth. However, the H$\alpha$ line profile matches \Hb\ well, notwirthstanding the contributions from the [\ion{N}{2}] narrow lines. This is a case where the SBHB hypothesis is disfavored.

\noindent \emph{J171448.--}
Matching the blue wing of \Hb\ to the Ly$\alpha$ and \ion{C}{4} lines implies that the red wings of the line profiles, including their peaks, are heavily absorbed. The shape of the UV line profiles suggests that this is not impossible, especially in the case of Ly$\alpha$ where the red wing of the line is absorbed by \ion{N}{5}. Notably, the \ion{Mg}{2} line is unabsorbed and appears consistent with \Hb. This is interesting, since \ion{Mg}{2} is a resonance line and therefore can show absorption, but it does not. The physical resolution is that this object has high-ionization absorption troughs indicating absorption in only high-ionization gas. Thus, this is a case for tentative support where absorption complicates the interpretation, but the UV line profiles are feasibly consistent with the SBHB hypothesis.

\section{Discussion}
\label{sec:discussion}
The approach of comparing the UV broad emission lines to the optical lines has proven to be an effective way to test the nature of SBHB candidates selected for radial velocity offset broad lines. It complements conclusions drawn from monitoring the radial velocity changes of the optical line profiles, since the perturbed accretion disk explanation can also produce similar drifts in peak velocity on multi-year timescales. Of the candidates where we obtained observations, we found support for the SBHB hypothesis in 3 candidates, tentative support in 5, disfavored 1, and found 4 where the absorption in the UV line profiles was too severe to draw any conclusion. 

This work has also revealed two limitations the approach of comparing the UV and optical line profiles, the first of which is the fact that the UV line profiles are frequently impacted by absorption that limits the comparison. We found 2/13 cases (15\%) where absorption introduced uncertainty into our conclusions, and an additional 4/13 (30\%) cases where the absorption was so severe that we were not able to draw any conclusions about the nature of the candidate. Such absorption lines superposed on the emission profiles is a common feature in AGN spectra \citep[e.g.,][]{ganguly08}, and arises from the fact that many prominent UV lines are subject to resonance scattering (absorption and isotropic re-emission of line photons) under the right conditions \citep{davidson79}. This includes \Lya, \CIV, and \MgII, although \MgII\ emission is produced by low-ionization dense gas \citep{collin-souffrin88} that is more likely to produce emission-line profiles similar to the optical Balmer lines \citep{halpern96}.

The second limitation to the UV spectroscopic test of the binary hypothesis is that there is some degree of subjectivity in classifying whether profiles are similar to each other or not. We aimed to minimize this by detailing the considerations that came into play for each source, and how they were prioritized in the decision making process (see \S\ref{sec:iobj}).

It is interesting to ask whether both red and blueshifted peaks are represented among the candidates that have evidence in support of the SBHB hypothesis. Notably, velocity offsets in both directions are expected in equal measure under the SBHB hypothesis and were represented among the candidates that were targeted with HST. However, outflows and winds, which are a common explanation for velocity offsets in the UV lines of quasars, produce only blueshifts \citep{richards11}. Among the eight candidates with tentative or strong support for the SBHB hypothesis, five have blueshifted broad lines. Absorption can also complicate the search for redshifted broad \Lya\ since \ion{N}{5} absorption is likely to impact a similar part of the spectrum, so these relative numbers are qualitatively consistent with expectations for a sample not driven by outflows.

Finally, the nature of this UV test is such that it is strongest when used in conjunction with other, complementary approaches to constraining the nature of the candidates. All of the candidates observed with HST for this work have radial velocity curves from \cite{eracleous12} or \citet{runnoe17}. In all cases, the radial velocity curves are consistent with the binary interpretation, although in most cases the curves are not yet well sampled so it is not the most stringent constraint. In five cases (J001224, J015530, J094603, J125142, J153636), limits have been placed on the physical properties that the binaries can have, but this did not rule any out \citep{runnoe17}. Ongoing optical spectroscopic monitoring will strengthen the radial velocity test for the candidates. Conversely, candidates with strong radial velocity credentials may arise that were not included in this HST sample due to the required integration time. It may be possible to conduct the UV spectroscopy test for these candidates, if (a) they were not included in the original sample and have short exposure times with HST, (b) the increased significance of their candidacy warrants longer HST exposure times, or (c) they are suitable targets for upcoming UV missions like the Ultraviolet Explorer \citep[UVEX;][]{kulkarni21}.

\section{Summary and Conclusions}
\label{sec:summary}
In this work, we performed a test of the nature of 13 sub-parsec separation SBHB candidates. The candidates were selected to have substantial ($\sim10^{3}$~\kms) velocity offsets between their broad \Hb\ emission lines and the rest frame set by [\ion{O}{3}], potentially the result of bulk orbital motion of one black hole and its bound BLR in a binary system. Here, we tested the alternative scenario that the offset broad Balmer lines and their time-dependent changes may be the result of a perturbed accretion disk and its dynamical evolution. We used HST observations of the broad UV line profiles, especially \Lya\ and \CIV, to discriminate between the SBHB and perturbed accretion disk scenarios. True SBHBs selected via the above approach are expected to show similar profiles and velocity offsets for all of the broad emission lines, so the observation of UV lines that are single peaked at the rest frame set by the narrow lines disfavors this scenario. In that case, a single AGN with non-axisymmetric BLR emission becomes the preferred explanation for the broad lines and their time-dependent changes.

This approach to testing the nature of SBHB candidates has proven to be effective and complementary to others already in use. We found strong support for the SBHB hypothesis in 3 candidates, tentative support in 5, 1 that was disfavored, and 4 cases where the absorption in the UV line profiles was too severe to conduct the test. From the basis of this work, the SBHB case for a number of candidates is significantly strengthened because they have passed an additional test of their nature.

\section*{Acknowledgements}

%
JCR and ME are grateful to Todd Boroson, for his fundamental contributions to the intellectual development of this project and the HST proposal that yielded the UV spectra used in this work. We also thank Gavin Mathes, who conducted some of the data reduction at the early stages of this project. As well, we acknowledge Tod Lauer and H{\'e}l{\`e}ne M.\,L.\,G. Flohic for their early contributions to the design of this project.

Support for program GO-12299 was provided by NASA through a grant from the Space Telescope Science Institute, which is operated by the Association of Universities for Research in Astronomy, Inc., under NASA contract NAS5-26555.

This work was partially supported by grant AST-1211756 from the National Science Foundation and an associated REU supplement. We thank the staff at Kitt Peak National Observatory, Apache Point Observatory, and the Hobby-Eberly Telescope for their expert help in carrying out the observations. 

JCR acknowledges support from the National National Foundation (NSF) from grant NSF AST-2205719 and the NASA Preparatory Science program under award 20-LPS20-0013.

TB acknowledges support from the NSF from grant NSF AST-2307278 and the Research Corporation for Science Advancement from award CS-SEED-2023-008.

The UV data presented in this paper were obtained from the Mikulski Archive for Space Telescopes (MAST) at the Space Telescope Science Institute. The specific observations analyzed can be accessed via \dataset[https://doi.org/10.17909/h3nm-jb62]{https://doi.org/10.17909/h3nm-jb62} STScI is operated by the Association of Universities for Research in Astronomy, Inc., under NASA contract NAS5–26555. Support to MAST for these data is provided by the NASA Office of Space Science via grant NAG5–7584 and by other grants and contracts.

We thank the staff at Apache Point Observatory, the Hobby-Eberly Telescope, Kitt Peak National Observatory, MDM Observatory, and Palomar Observatory, for their expert help in carrying out the observations. 

Funding for the SDSS and SDSS-II has been provided by the Alfred P. Sloan Foundation, the Participating Institutions, the National Science Foundation, the U.S. Department of Energy, the National Aeronautics and Space Administration, the Japanese Monbukagakusho, the Max Planck Society, and the Higher Education Funding Council for England. The SDSS Web Site is http://www.sdss.org/.

The SDSS is managed by the Astrophysical Research Consortium for the Participating Institutions. The Participating Institutions are the American Museum of Natural History, Astrophysical Institute Potsdam, University of Basel, University of Cambridge, Case Western Reserve University, University of Chicago, Drexel University, Fermilab, the Institute for Advanced Study, the Japan Participation Group, Johns Hopkins University, the Joint Institute for Nuclear Astrophysics, the Kavli Institute for Particle Astrophysics and Cosmology, the Korean Scientist Group, the Chinese Academy of Sciences (LAMOST), Los Alamos National Laboratory, the Max-Planck-Institute for Astronomy (MPIA), the Max-Planck-Institute for Astrophysics (MPA), New Mexico State University, Ohio State University, University of Pittsburgh, University of Portsmouth, Princeton University, the United States Naval Observatory, and the University of Washington.

This work is based on observations at Kitt Peak National Observatory, which is operated by the Association of Universities for Research in Astronomy (AURA) under cooperative agreement with the National Science Foundation.

This work is based on observations obtained with the Apache Point Observatory 3.5-meter telescope, which is owned and operated by the Astrophysical Research Consortium. 

This work is based on observations obtained at the MDM Observatory, operated by Dartmouth College, Columbia University, Ohio State University, Ohio University, and the University of Michigan.



The Marcario Low-Resolution Spectrograph is named for Mike Marcario of High Lonesome Optics, who fabricated several optics for the instrument but died before its completion; it is a joint project of the Hobby-Eberly Telescope partnership and the Instituto de Astronom\'{\i}a de la Universidad Nacional Aut\'onoma de M\'exico.


%

\facilities{HST(COS, STIS), Hobby-Eberly Telescope, Apache Point Observatory ARC 3.5m, Kitt Peak Observatory Mayall 4m, MDM Observatory, Palomar Observatory}


\software{stistools \citep{stistools},
          pyspeckit \citep{ginsburg22},
          astropy \citep{astropy13,astropy18},
          IRAF \citep{iraf}
          }




\bibliography{references}{}
\bibliographystyle{aasjournal}



\end{document}